\documentclass[aps, pra, superscriptaddress, floatfix, reprint]{revtex4-1}
\usepackage{graphicx}
\usepackage{dcolumn}
\usepackage{bm}
\usepackage{amsfonts}
\usepackage{amsmath}
\usepackage{amssymb}
\usepackage{color}
\usepackage[normalem]{ulem}
\usepackage[percent]{overpic}

\usepackage[colorlinks=true,citecolor=blue]{hyperref}
\hypersetup{colorlinks=true,citecolor=blue,linkcolor=red,urlcolor=blue}

\def\be{\begin{equation}}
\def\ee{\end{equation}}

\begin{document}

\title{Rydberg-dressed Fermi liquid: correlations and signatures of droplet crystallization}
\author{Iran Seydi}
	\affiliation{Department of Physics, Institute for Advanced Studies in Basic Sciences (IASBS), Zanjan 45137-66731, Iran}
\author{Saeed H. Abedinpour}
\email{abedinpour@iasbs.ac.ir}
	\affiliation{Department of Physics, Institute for Advanced Studies in Basic Sciences (IASBS), Zanjan 45137-66731, Iran}
	\affiliation{Research Center for Basic Sciences \& Modern Technologies (RBST), Institute for Advanced Studies in Basic Sciences (IASBS), Zanjan 45137-66731, Iran}
		\affiliation{School of Nano Science, Institute for Research in Fundamental Sciences (IPM), Tehran 19395-5531, Iran}
\author{Reza Asgari}
\affiliation{School of Physics, Institute for Research in Fundamental Sciences (IPM), Tehran 19395-5531, Iran}
\author{Martin Panholzer}
\affiliation{Institute for Theoretical Physics, Johannes Kepler University, Altenbergerstrasse 69, 4040 Linz, Austria}
\affiliation{Uni Software Plus GmbH, 4320 Perg, Austria}
\author{B. Tanatar}
	\affiliation{Department of Physics, Bilkent University, Bilkent, 06800 Ankara, Turkey}

\date{\today}

\date{\today}
\begin{abstract}
We investigate the effects of many-body correlations on the ground-state properties of a single component ultra-cold Rydberg-dressed Fermi liquid with purely repulsive inter-particle interactions, in both three and two spatial dimensions.
We have employed the Fermi-hypernetted-chain Euler-Lagrange approximation and observed that the contribution of the correlation energy on the ground-state energy becomes significant at intermediate values of the soft-core radius and large coupling strengths. For small and large soft-core radii, the correlation energy is negligible and the ground-state energy approaches the Hartree-Fock value. 
The positions of the main peaks in static structure factor and pair distribution function in the homogeneous fluid phase signal the formation of quantum droplet crystals with several particles confined inside each droplet.
\end{abstract}
\maketitle

\section{Introduction}\label{sect:intro}
Ultra-cold atoms can provide clean and controllable experimental tools to explore novel quantum phases of matter. These systems enjoy an artificial inter-particle interaction that usually does not have a counterpart in conventional condensed matter systems. 
Rydberg atom systems, due to their long-range and strong dipole-dipole interactions, are suitable for constructing strongly correlated models and for many-body simulations~\cite{bernien2017, zeiher2016many, labuhn2016tunable,balewski2014NewJournalofPhysics}.
Rydberg atoms have many applications in non-linear quantum optics
~\cite{firstenberg2016nonlinear, dudin2012strongly,Adams_2019}
, quantum information~\cite{saffman2010quantum, isenhower2010demonstration, lukin2001dipole, browaeys2016interacting}, quantum simulation~\cite{weimer2010rydberg}, and in the study of biophysical transport phenomena~\cite{plodzien2018scientificreports}.
Optical imaging of the shape of electron orbitals of neutral atoms in Bose-Einstein condensation of Rydberg atoms has been proposed by Karpiuk \emph{et al.}~\cite{karpiuk2015imaging}.

 Usually, the lifetime of a Rydberg state is not long enough to allow the study of the atomic dynamics, but to enhance the lifetime of Rydberg atoms, the ground state could be coupled to the Rydberg state off-resonantly. In other words, with a small admixture of the Rydberg state in the ground state, it is possible to obtain long-lived Rydberg-dressed states~\cite{henkel2010three, browaeys2016experimental, zeiher2016many}. 
Rydberg dressing for two atoms~\cite{jau2016entangling} and in optical lattices~\cite{zeiher2017coherent, zeiher2016many}, has been observed experimentally. The microscopy of Rydberg macro-dimers has been reported as well~\cite{Hollerith2019}. 
Rydberg-dressed atoms can be employed in the search for novel quantum phases such as the super-solid phase~\cite{henkel2010three, pupillo2010strongly, boninsegni2012colloquium, PhysRevLett.105.135301, Cinti,prestipino2018prb}, quantum liquid droplets~\cite{Cinti,PhysRevLett.120.235301,PhysRevLett.116.215301,cabrera2018quantum,Seydi2019arxiv,PhysRevLett.116.215301},  topological quantum magnetism~\cite{PhysRevLett.110.257204}, topological superfluidity~\cite{xiong2014topological}, mixed topological density-wave~\cite{li2015exotic}, and quantum spin-ice~\cite{PhysRevX.4.041037}. 

The long-anticipated super-solid phase has been finally observed in ultra-cold dipolar systems of magnetic atoms, very recently~\cite{tanzi2019prl, bottcher2019transient, chomaz2019long}. Therefore, the observation of the predicted super-solid or droplet solid phases in Rydberg-dressed systems appears very feasible. 

The effect of inter-particle interactions in a three-dimensional (3D) Rydberg-dressed Fermi system with a pure repulsive interaction has been studied within the mean-field approximation and the density-wave instability to a metallic quantum solid phase has been predicted~\cite{li2016emergence}. 
The ground-state properties of a two-dimensional (2D) Rydberg-dressed Fermi liquid has been investigated in the framework of the Hartree-Fock approximation and functional renormalization group~\cite{khasseh2017phase, keles2019phase}. 
The density-wave instability of the homogenous system has been reported using the random-phase approximation (RPA)~\cite{khasseh2017phase}. With the help of the functional renormalization group method, both f-wave superfluidity and density-wave instability has been predicted for Rydberg-dressed fermions with repulsive interaction in 2D~\cite{keles2019phase}. 

In this work, we address the effects of many-body correlation on the ground state properties of a single component Rydberg-dressed Fermi liquid in both three- and two-dimensions, within the Fermi hyper-netted chain Euler-Lagrange (FHNC-EL) formalism at zero temperature. 
We show that the impact of the correlation energy on the ground-state energy becomes significant only at intermediate values of the soft-core radius and large coupling strengths.  
Having calculated the positions of the main peaks in the static structure
factor and the pair distribution functions in the homogeneous fluid phase, we anticipate instability of the homogeneous fluid to form quantum droplet crystals with several particles confined inside each droplet. In the absence of any exact or state of the art quantum Monte-Carlo (QMC) simulation results for Rydberg-dressed fermions, we aim to verify the validity domain of the mean-field approximations as well the regimes of the system parameters where the beyond mean-field effects become significant. The FHNC-EL formalism has been shown to provide a very accurate account of the many-body correlations in the homogeneous liquid phase~\cite{PhysRevB.68.155112, ASGARI2004301, ABEDINPOUR201425}, with orders of magnitude less computational demand in comparison to the QMC simulations. Furthermore, the analytic treatment of the ground state in FHNC methods allows an extension to dynamic properties~\cite{PhysRevB.82.224505}.

The rest of this paper is organized as follows. 
In Sec.~\ref{sec:theory}, we describe our theoretical formalism and review the details of the FHNC-EL approximation. 
In Sec.~\ref{sec:result}, we present our numerical results for different ground-state quantities of the homogeneous fluid phase such as the static structure factor, pair distribution function, effective interaction, and the ground state energy. Furthermore, we investigate the instability of the homogeneous phase towards density modulated phases. Finally, Sec.~\ref{sec:sum} summarizes our main findings.

\section{Model and theoretical formalism}\label{sec:theory}
We consider a single component gas of Rydberg-dressed fermions with the average density of $n$, and the bare mass of particles $m$, confined in a two- or three-dimensional space. The interaction between two Rydberg-dressed atoms is given by~\cite{henkel2010three}
\begin{equation}\label{eq:Rydberg int}
v_{\rm RD}(r)= \frac{U}{1 + (r/R_c)^6},
\end{equation}
where $R_{c}  = [C_{6} / (2 \hbar \Delta)]^{1/6}$ is the soft-core radius of interaction and $U = [\Omega/(2\Delta)]^{4} C_{6} / R_{c}^{6}$ is the interaction strength. Here $\Omega$, $\Delta < 0$, and $C_{6} < 0$ are the effective Raman coupling, red detuning and averaged van der Waals coefficient, respectively.  
The ground-state properties of this gas could be characterized in terms of two dimensionless parameters
$u = U / \varepsilon_{\rm F}$ and $r_c = R_{c} k_{\rm F}$, where 
$\varepsilon_{\rm F} = {\hbar}^{2} k_{\rm F}^{2}/(2m)$ is the Fermi energy and $k_{\rm F} = (2 {\rm d} \pi^{{\rm d}-1}n)^{1/{\rm d}}$ is the Fermi wave vector in d-spatial dimensions with d=2 or 3.

\subsection{Fermi-hypernetted-chain Euler-Lagrange approximation}\label{sec:FHNC}
Taking the chemical potential of the system as the zero point of energy, we can write a formally exact differential equation for the pair distribution function $g(r)$~\cite{PhysRevB.68.155112, ASGARI2004301}
\begin{equation}\label{eq:diff_g}
\bigg[ -\frac{\hbar^2}{m} \nabla_{\textbf{r}}^{2} + w_{\rm eff}(r)\bigg]\sqrt{g(r)} = 0,
\end{equation}
where $w_{\rm eff}(r)$ is the effective potential, and in practice needs to be approximated. 
Unlike the bosonic systems, a truly FHNC formulation for the effective interaction in Fermi gases leads to a very complicated set of coupled equations~\cite{Krotscheck2000}. 
However, several simplified recipes have tried to implement the exact weak or strong coupling limiting behaviors in the effective interaction and proved to give reasonably accurate results in the corresponding limits~\cite{kallio1996novel,panholzer2018optimized}. 
In this work, we follow the recipe of Kallio and Piilo (KP), which has been tailored to exactly capture the Fermi statistics and weak coupling behavior~\cite{kallio1996novel}. For an alternative approximation, based on the approximate summation of ladder and ring diagrams see Appendix~\ref{sec:panholzer}.

Within the KP approximation the effective interaction is given by
\begin{equation}\label{eq:eff_V}
w_{\rm eff}(r) = v_{\rm RD}(r) + w_{\rm B}(r) + w_{\rm F}(r),
\end{equation}
where the bosonic potential $w_{\rm B}(q)$ in the Fourier space, at the level of HNC-EL/0 approximation, \emph{i.e.}, neglecting the elementary diagrams and correlations higher than pair correlations, is given by
\be\label{eq:W_B}	
\begin{split}
w_{\rm B}(q) &= -\frac{\varepsilon_q}{2n} \left[2S(q) + 1\right] \left[\frac{S(q) - 1}{S(q)}\right]^{2}. 
\end{split}
\ee
Here, $\varepsilon_q=\hbar^2 q^2/(2 m)$ is the non-interacting dispersion and $S(q)$ is the static structure factor, related to the pair distribution function as $S(q) = 1 + n {\rm FT}[g(r) - 1]$, where the Fourier transform (FT) is defined as: 
$\int \mathrm{d}{\textbf {r}} f(r) e^{-i\textbf{q}\cdot\textbf{r}}$.
The Fermi contribution to the effective potential $w_{\rm F}(r)$, which includes most importantly the exchange effects, within the KP approximation reads
\be\label{eq:W_F}
w_{\rm F}(r) = \frac{\hbar^2}{m} \frac{\nabla_{\textbf{r}}^{2} \sqrt{g_{0}(r)}}{\sqrt{g_{0}(r)}} - \lim_{u \rightarrow 0} w_{\rm B}(r),
\ee
where $g_{0}(r)$ is the pair distribution function of a non-interacting Fermi gas~\cite{giuliani2005quantum} and the non-interacting limit of the Bose potential, could be obtained after replacing the static structure factor in Eq.~\eqref{eq:W_B} with  $S_0(q)$, the static structure factor of a non-interacting Fermi gas~\cite{giuliani2005quantum}.

A numerically efficient strategy to solve the zero-energy differential equation~\eqref{eq:diff_g}, is to invert it and obtain the effective potential
\begin{equation}\label{eq: V_ph}
V(r) = g(r) w_{\rm eff}(r)- w_{\rm B}(r) + \frac{\hbar^2}{m} {\arrowvert \nabla \sqrt{g(r)}}\arrowvert^2,
\end{equation}
whose Fourier transform gives the static structure factor
\begin{equation}\label{eq:static S}
S(q) = \frac{1}{\sqrt{1 + 2 n V(q) / \epsilon_q}}.
\end{equation} 
Now, Eqs.\,\eqref{eq:W_B}, ~\eqref{eq: V_ph} and ~\eqref{eq:static S} could be solved self-consistently for a given set of system parameters $u$ and $r_c$. Note that the Fermi potential in the KP formalism, as given by Eq.\,\eqref{eq:W_F} is already fixed by the non-interacting pair distribution function and structure factor, and does not enter the loop of self consistency.

\subsection{The ground-state energy}
Once the pair distribution function is known for different interaction strengths, the ground-state energy per particle could be obtained from the coupling constant integration~\cite{giuliani2005quantum}
\be\label{eq:e_coupling}
\varepsilon_{\rm GS}(u,r_c) 
 = \varepsilon_0 + \frac{n}{2} \int\limits_{0}^{u} \mathrm{d} u' \int \mathrm{d} \textbf{r} \, \frac{g_{u'}(r)}{1 + (r/R_c)^{6}}. 
\ee 
Here, $ \varepsilon_0=\varepsilon_{\rm F} {\rm d} /({\rm d}+2)$ is the non-interacting kinetic energy of a ${\rm d}$-dimensional Fermi gas and $g_{u'}(r)$ is the interacting pair distribution function of a Rydberg-dressed Fermi liquid with interaction strength equal to $u'$ and at fixed soft core radius $r_c$.
The correlation energy, which is defined as the difference between the exact ground-state and the restricted Hartree-Fock energies is a good measure of the performance of any approximate theories.  In the next section, we will report our numerical results for the ground-state and correlation energies as functions of the interaction strength $u$, and the soft-core radius $r_c$. 

\subsection{Density-density response function}
The collective density modes and signatures of the instability of a homogeneous system to density modulated phases both could be obtained from the singularities of its density-density response function
\begin{equation}\label{eq:chi_qw}
\chi (q , \omega) = \frac{\chi_{0}(q , \omega)}{1 - V_{\rm ph} (q,\omega) \chi_{0}(q , \omega) },
\end{equation}
respectively in the dynamic and static regimes. 
Here, $\chi_{0}(q , \omega)$ is the non-interacting density-density response function~\cite{giuliani2005quantum}, and 
$V_{\rm ph} (q,\omega)$ is the particle-hole irreducible interaction~\cite{PhysRevB.82.224505}, which  needs to be approximated for any practical purpose. 
In the acclaimed random-phase approximation (RPA), all the exchange and correlation effects are discarded replacing the effective interaction with the bare interaction. 
Schemes to go beyond the RPA mainly rely on introducing the many-body local-field factors~\cite{giuliani2005quantum}.
On the other hand, if the interacting static structure factor is known, the fluctuation-dissipation theorem
\begin{equation}\label{eq:FDT}
S(q) = -\frac{\hbar}{n \pi} \int_{0}^{\infty} \mathrm d \omega \,\Im m [\chi(q , \omega)],
\end{equation}
could be used to extract a static effective interaction~\cite{ABEDINPOUR201425}. 
Further approximating the non-interacting density-density response function of the Fermi gas with a Bose-like expression, \emph{i.e.}, the ``mean spherical approximation" (MSA)
\begin{equation}
\chi_{0}^{\rm MSA}(q, \omega) = \frac{2n \varepsilon_{q}}{\left(\hbar \omega + i 0^+\right)^{2} -\left(\varepsilon_{q} / S_{0}(q)\right)^{2}},
\end{equation}
the frequency integral in Eq.\,\eqref{eq:FDT} could be performed analytically and a simple analytic expression for the static effective interaction is obtained
\begin{equation}\label{eq:psi_eff}
V_{\rm ph}(q) = \frac{\varepsilon_q}{2n}\left[\frac{1}{S^{2}(q)} - \frac{1}{S_{0}^{2}(q)}\right].
\end{equation}
In this work, we use the static structure factor obtained from the solution of KP equations to extract the static effective interaction. This approach has proved to give very good results for different properties of various strongly interacting Fermi liquids~\cite{Seydi_PRA2018, asgari2006many}.

\section{Numerical Results and Discussion}\label{sec:result}
In this section, we turn to the presentation of our numerical results for static structure factor, pair distribution function, and effective interaction $w_{\rm eff}(r)$ of a one-component Rydberg-dressed Fermi liquid obtained from the KP approximation. 
We also investigate the contribution of correlation energy to the total ground-state energy at different system parameters. Finally, we discuss the dynamical structure factor and the density-wave instability of the homogeneous Rydberg-dressed Fermi liquid.

\subsection{Static structure factor}\label{sec:S_q}
\begin{figure}
	\includegraphics[width=0.5\textwidth]{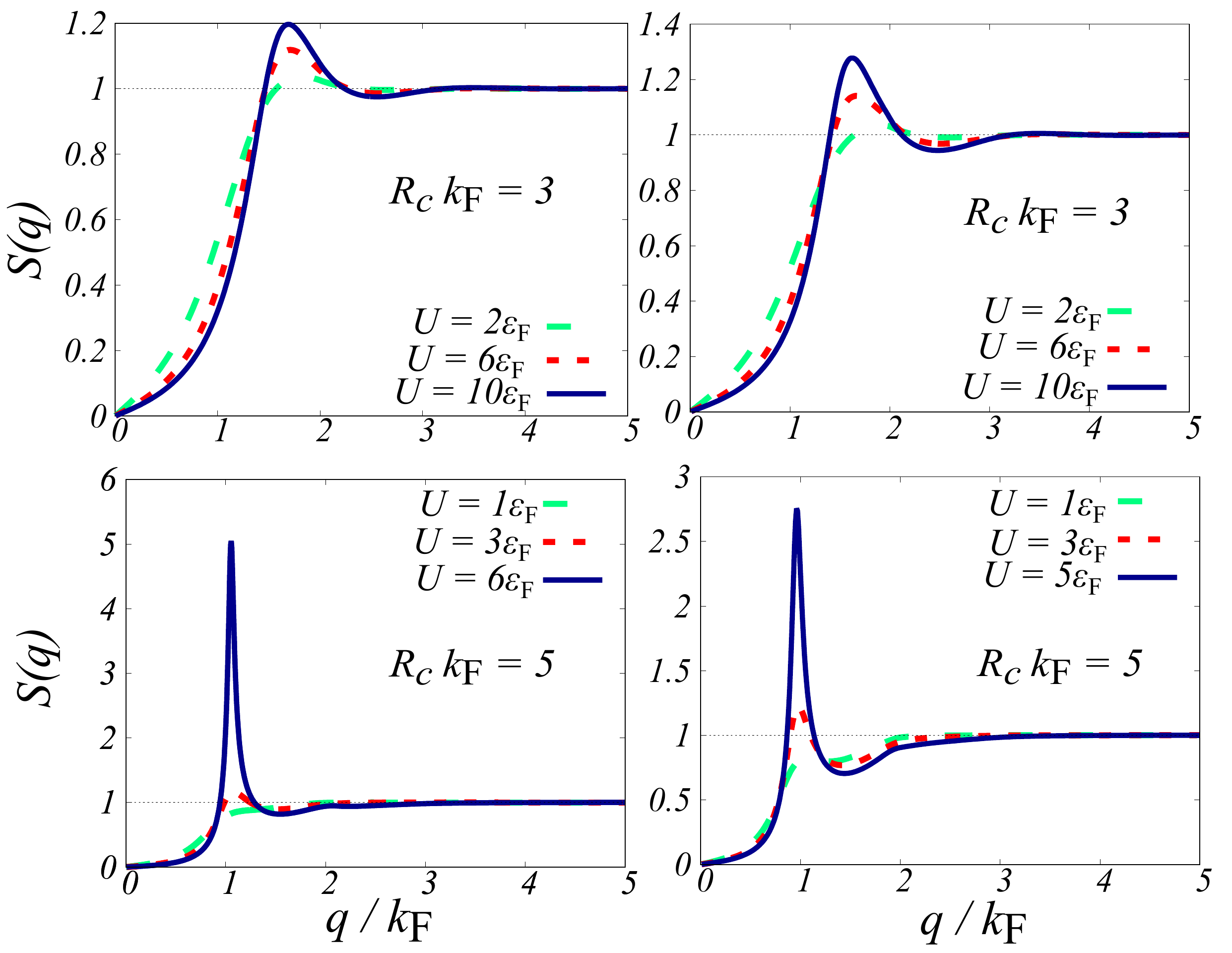}	
	\caption{The static structure factor of 3D (left) and 2D (right) Rydberg-dressed Fermi liquid versus $q / k_{\rm F}$ obtained from the solution of KP equations at two fixed values of the dimensionless soft-core radius $r_c = R_c k_F$ and for different values of the interaction strength $u=U/\varepsilon_{\rm F}$.  \label{fig:Sq}}
\end{figure}

Figure\,\ref{fig:Sq} shows the static structure factors of three- and two-dimensional Rydberg-dressed Fermi liquids, obtained from the solution of KP equations at different values of the soft-core radius $r_c$ and coupling 
constant $u$.
For a fixed value of the soft-core radius, increasing the interaction strength, the height of the main peak in $S(q)$ increases, indicating the enhancement of correlations.
Another interesting observation in the behavior of the static structure factor is the location of its main peak. 
This is usually fixed by the density of system \emph{i.e.}, the average spacing between particles and is related to the position of Bragg peak in the crystalline phase.  
Our numerical results show that the position of this main peak for Rydberg-dressed atoms shifts to smaller wave vectors as the correlations get stronger. This is an indication of the tendency of the system to the formation of crystalline structures with lattice constants larger than the average inter-particle spacing.

\subsection{Pair distribution function and effective interaction}\label{sec:gr}
The pair distribution function $g(r)$ gives the relative spatial distribution of particles in the system, therefore is a positive-definite function.
The pair distribution function is normalized such that  
 $\lim \limits_{r \rightarrow \infty} g(r) \rightarrow 1$, since the correlation between particles vanishes at large separations.
 
\begin{figure}
	\includegraphics[width=0.5\textwidth]{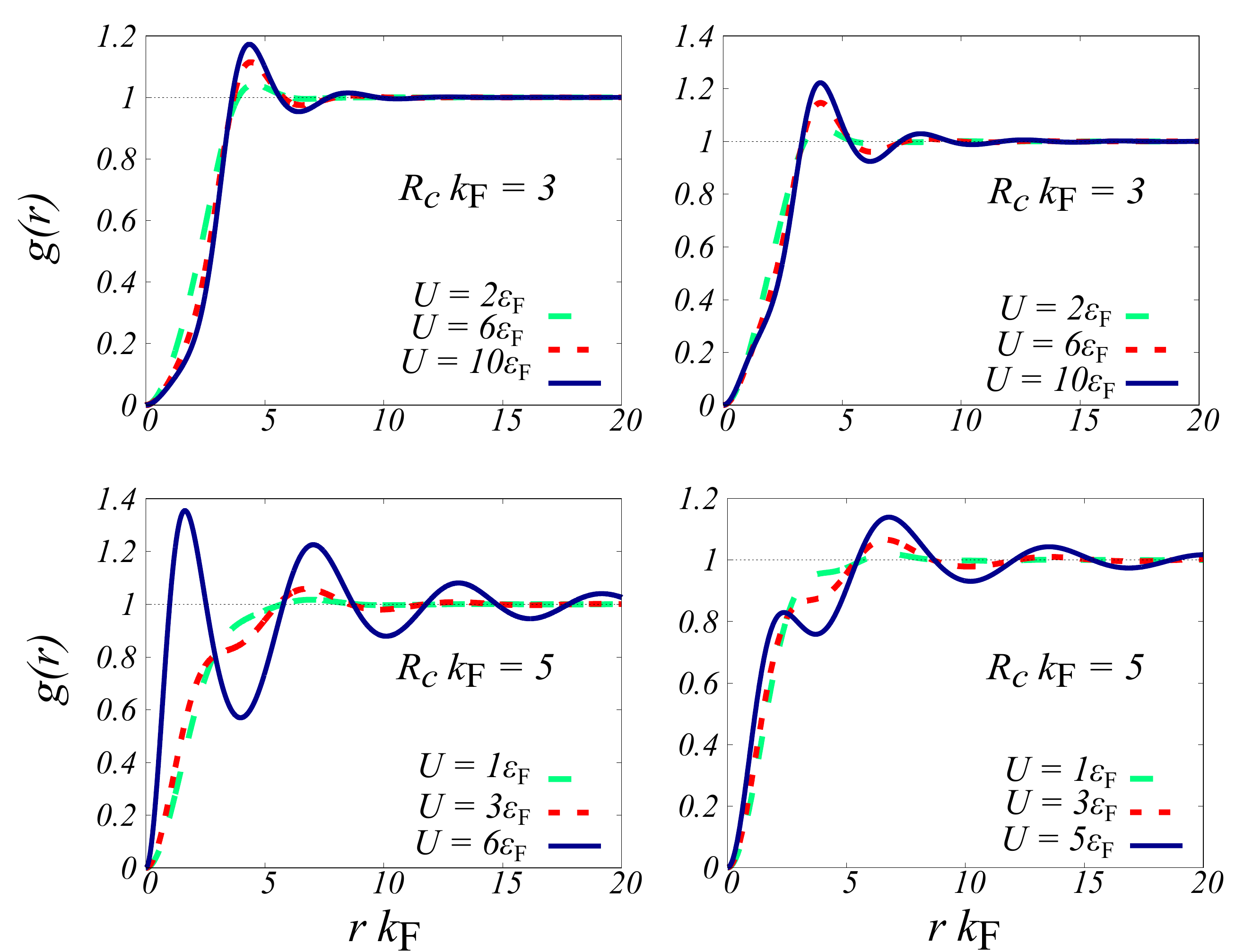}	
	\caption{The pair distribution function versus $r k_{\rm F}$ within the KP approximation at two fixed values of the soft-core radius and for different values of the interaction strength for 3D (left) and 2D (right) Rydberg-dressed Fermi liquids.  \label{fig:gr}} 
\end{figure}

Figure\,\ref{fig:gr} illustrates the pair distribution function of Rydberg-dressed Fermi liquids at different values of the soft-core radius $r_c$ and coupling constant $u$. 
It is interesting to note that the positivity of the pair distribution function and the exact condition $g (r = 0)= 0$ for our spin-polarized Fermi system, as implied by Pauli's exclusion principle, are both satisfied within the approximate KP formalism, even up to very strong interaction strengths.

Apart from the well-expected behavior of the pair distribution function of the liquid phase at strong correlations, \emph{i.e.}, a pronounced first peak at a specific distance and slowly decaying oscillations, here an interesting observation is the appearance of a shoulder at small distances which evolves into a peak and eventually dominates the original first-peak of the pair distribution function at large soft core radius and strong couplings. This indicates a smaller first-neighbor distance than the average inter-particle separation or the tendency of particles to aggregate at strong interactions. This is quite counterintuitive, keeping in mind the repulsive nature of the bare inter-particle interaction. 
When considered together with the peculiar behavior of the main peak in the static structure factor, this could be an indication of the tendency of the system for the formation of quantum droplet crystals at strong correlations.
The average distance between ordered droplets becomes larger featured as the smaller wave vector peak in the static structure factor. At the same time, the first neighbor distance gets smaller due to the clustering of several particles inside each droplet. 

\begin{figure}
\includegraphics[width=0.5\textwidth]{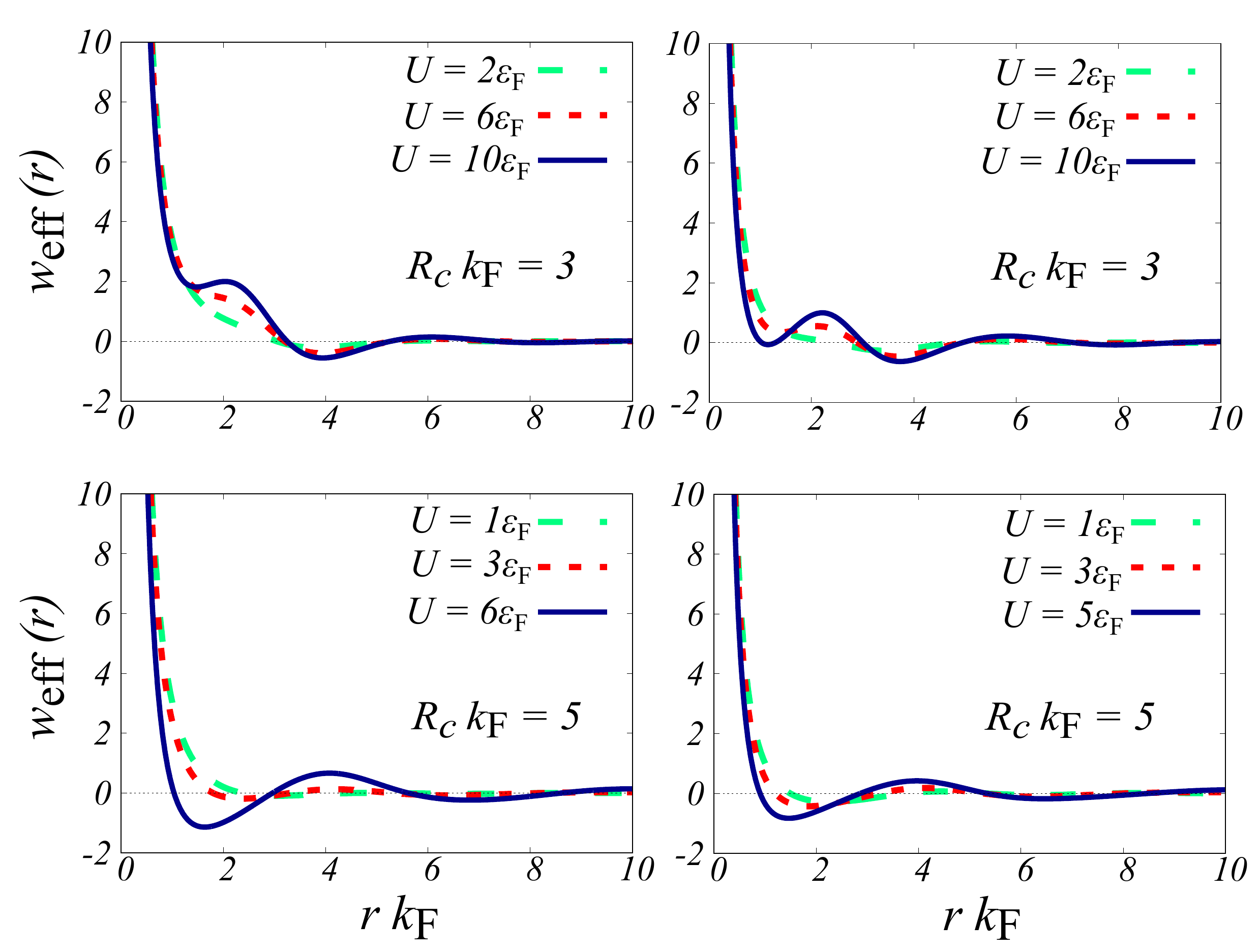}	
\caption{The effective interaction $w_{\rm eff}(r)$ (in units of the Fermi energy $\varepsilon_{\rm F}$) versus $r k_F$  obtained within the KP approximation at two fixed values of $R_{c} k_{\rm F}$ and for different values of the interaction strength for 3D (left) and 2D (right) Rydberg-dressed Fermi liquids.  
\label{fig:Weff}} 
\end{figure}

This distinctive behavior of the pair distribution function could be understood from the effective interaction $w_{\rm eff}(r)$, 
as illustrated in Fig.\,\ref{fig:Weff}. 
Apart from the repulsive hard-core of the effective interaction which originates from the statistical \emph{i.e.}, Pauli repulsion, the effective interaction becomes attractive around the distance where the first peak of the pair distribution function appears. 
At strong couplings, the effective interaction has an oscillatory behavior, and its first minimum moves towards smaller distances, in agreement with the behavior of the pair distribution function. It is worth mentioning that quantum fluctuations are enhanced in 2D and the possibility of observing droplets in 2D is of great interest and as shown in Fig.\,\ref{fig:Weff}, attractive effective potential permits the formation of self-bound quantum droplets. 

\subsection{The ground-state and correlation energies}\label{sec:E_GS}
The ground-state energy per particle of the Rydberg-dressed Fermi liquid could be written in terms of different contributions to it
\be\label{eq:e_gs}
\varepsilon_{\rm GS}(u,r_c)=\varepsilon_0+\varepsilon_{\rm H}(u,r_c)+\varepsilon_{x}(u,r_c)+\varepsilon_{c}(u,r_c),
\ee
where $ \varepsilon_0=\varepsilon_{\rm F} d /(d+2)$ is the non-interacting kinetic energy of a $d$-dimensional Fermi gas and the Hartree energy is given by
\be\label{eq:e_H}
\varepsilon_{\rm H}(u, r_c) = \frac{n}{2} v_{\rm RD}(q = 0) =\alpha_d  u r_c^d  \varepsilon_{\rm F},
\ee
with $\alpha_d=(\pi/\sqrt{3})^{3-d}/(6 d)$.
The exchange energy per particle could be obtained from~\cite{giuliani2005quantum}
\begin{equation}\label{eq:e_x}
\varepsilon_{x}(u,r_c) 
= -\frac{1}{2NL^d}\sum_{\textbf q}v_{\rm RD}(q)\sum_{\textbf k} n_{\textbf k + \textbf q } n_{\textbf k}
\end{equation}
where $N$ is the particle number, $L$ is the sample length, and $n_{\bf k}$ is the Fermi-Dirac distribution function, which becomes a step function $\Theta(\varepsilon_{\rm F} - \varepsilon_k)$ at zero temperature.
Analytic summation over ${\bf q}$ and ${\bf k}$ in Eq.\,\eqref{eq:e_x} is possible and it yields
\be
\begin{split}
\varepsilon^{(\rm 3D)}_{x}(u , r_c) = \frac{\varepsilon_{\rm F} u}{4 r_c^3} 
\Bigg\{ &  e^{-2  r_c}\left(1 +   r_c\right)^2 -2  r_c^3 + 2 r_c^2  + 1  \\
&-e^{- r_c} \bigg[  \sqrt{3} \left( 2 r_c + r_c^2\right) \sin \left(\sqrt{3}r_c\right)  \\
& ~~+ \left( 2 + 2  r_c - r_c^2\right) \cos \left(\sqrt{3} r_c\right)\bigg]   \Bigg\}  ,
\end{split}
\ee
and~\cite{ khasseh2017phase}
\begin{equation}
\varepsilon_{x}^{(\rm 2D)} (u, r_c) = \frac{ \varepsilon_{\rm F} u}{2}
\left[ 1 -
\frac{G^{4,3}_{3,9}\left(\frac{r_c^6}{3^6} \left|_{0,\frac{1}{3},1,-\frac{1}{3}, \frac{1}{3}, \frac{2}{3}}^{\frac{1}{6}, \frac{1}{2}, \frac{5}{6}}\right.\right)}{3\sqrt{3}} \right],
\end{equation}
in three- and two-dimensions, respectively. Here, $G(\cdots)$ is a Meier G-function~\cite{gradshteyn1996table}.

The correlation energy $\varepsilon_c$ itself is defined through Eq.\,\eqref{eq:e_gs} as the difference between the exact ground-state energy and the Hartree-Fock energy, and needs to be approximated. We use the coupling constant integration as introduced through Eq.\,\eqref{eq:e_coupling}, with the pair distribution function obtained from the KP calculations to find the correlation energy of three- and two-dimensional Rydberg-dressed Fermi liquids for different values of the coupling strengths and soft-core radii.

\begin{figure}
\includegraphics[width=0.5\textwidth]{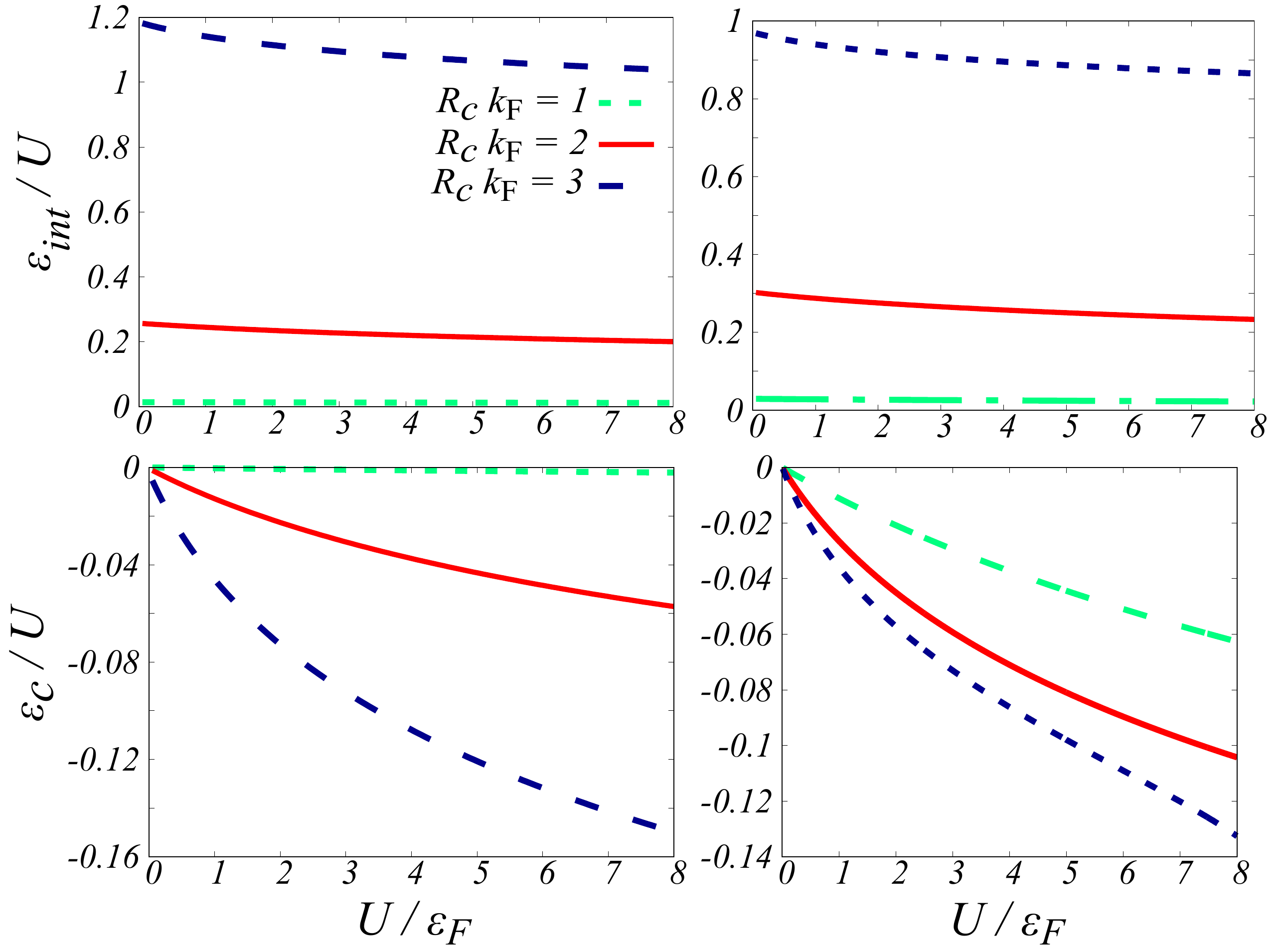}
\caption{Top panels: the interacting part of the ground-state energy per-particle $\varepsilon_{\rm int}$  (in units of the interaction strength $U$) versus $U/\varepsilon_{\rm F}$ for different values of the soft-core radius, obtained from the KP approximation for 3D (left) and 2D (right) Rydberg-dressed Fermi liquids.
Bottom panels: same as the top panels, but for the correlation energy $\varepsilon_c$.
 \label{fig:E_GS}}
\end{figure}

In Fig.\,\ref{fig:E_GS} we present the interaction contribution to the ground-state energy $\varepsilon_{\rm int}=\varepsilon_{\rm GS}-\varepsilon_{0}$ and the correlation energy $\varepsilon_c$, versus the interaction strength for different values of the soft-core radius. The correlation energy, as expected, is negative but the total ground-state energy is an increasing function of both $u$ and $r_c$. As it is clear from the analytic expressions~\eqref{eq:e_H} and~\eqref{eq:e_x}, very weak non-linear $u$ dependence of the interaction energy (top panels) originates solely from the correlation energy (bottom panels). 

\begin{figure}
\includegraphics[width=0.5\textwidth]{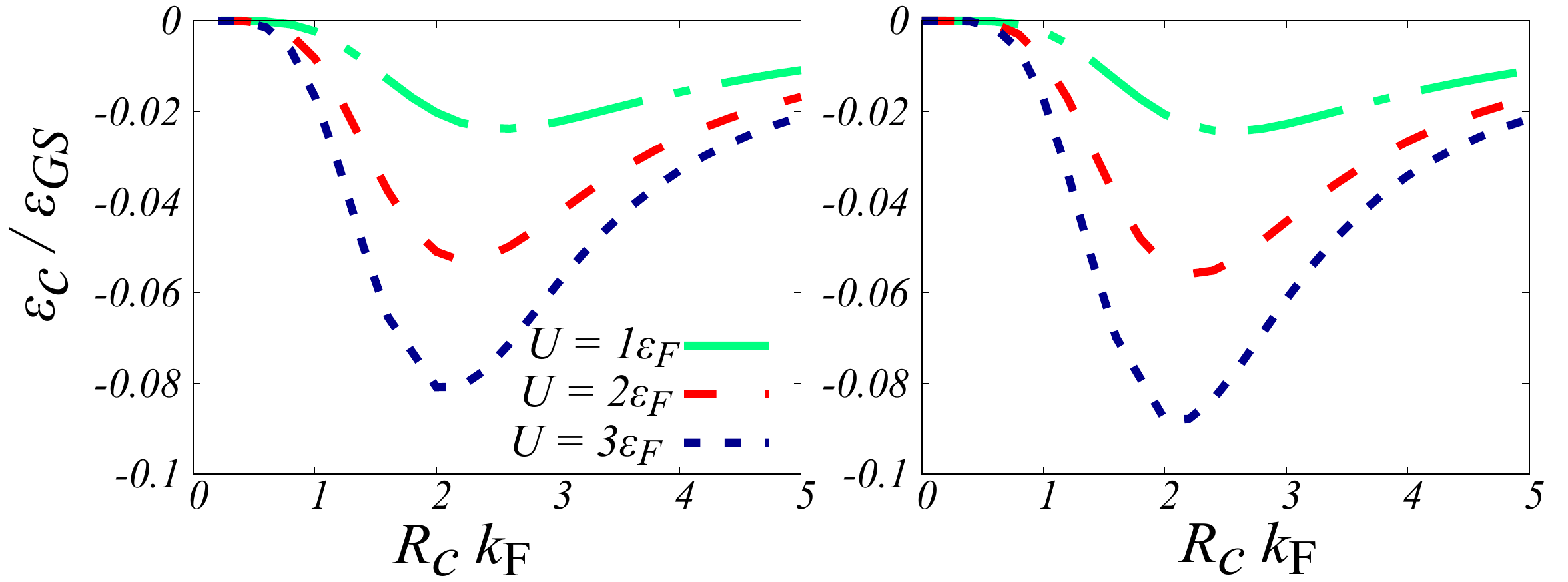} 
	\caption{The correlation energy per particle $\varepsilon_c$ in units of the total ground-state energy
		$\varepsilon_{\rm GS}$, as a function of the soft-core radius $r_c$ at several values
		of the coupling strength $u$, calculated within the KP formalism for 3D (left)
		and 2D (right) Rydberg-dressed Fermi liquids.
		\label{fig:Ec}}	
\end{figure}

Figure\,\ref{fig:Ec} illustrates the soft-core radius dependence of the correlation energy of a Rydberg-dressed Fermi liquid. The correlation energy has a considerable contribution to the ground-state energy only at intermediate values of the soft-core radii (\emph{i.e.}, $r_c \approx 2$) and at large coupling strengths. For both small and large values of $r_c$,
the correlation energy has a small value, and the KP ground-state energies approach the mean-field HF results. This can be attributed to the behavior of the bare potential. At small values of $R_c$, the potential is short-ranged decaying as $1/r^6$, and for large $R_c$ the potential is almost constant.

\subsection{The dynamical structure factor and collective modes}\label{sec:S_kw}
\begin{figure}
	\begin{tabular}{cc}
	\includegraphics[width=0.25\textwidth]{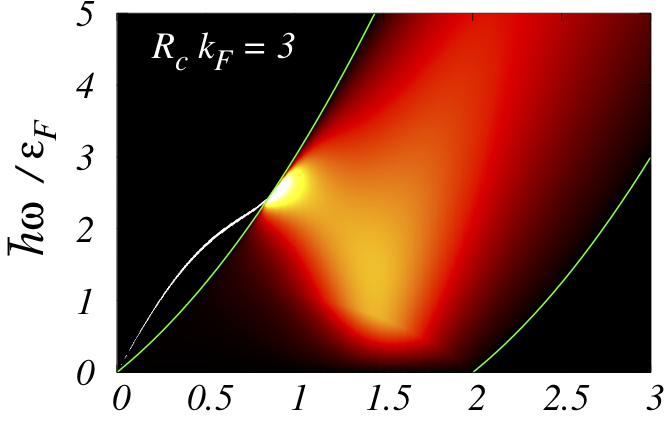}&
	\includegraphics[width=0.25\textwidth]{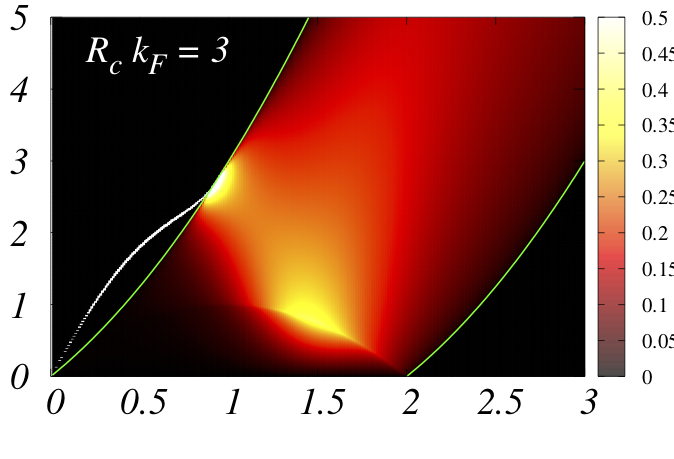}\\
	\includegraphics[width=0.25\textwidth]{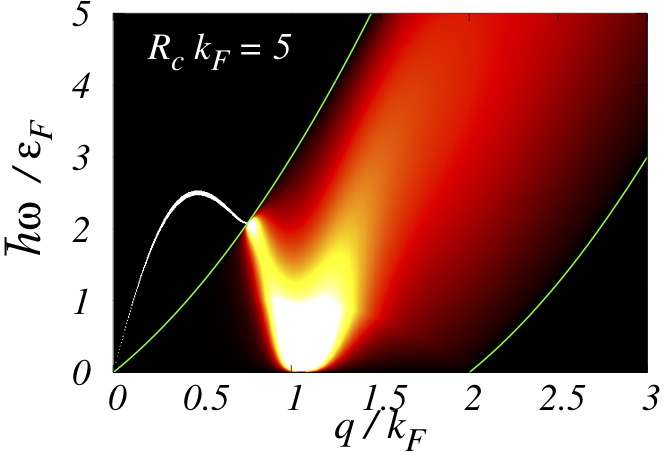}&
	\includegraphics[width=0.25\textwidth]{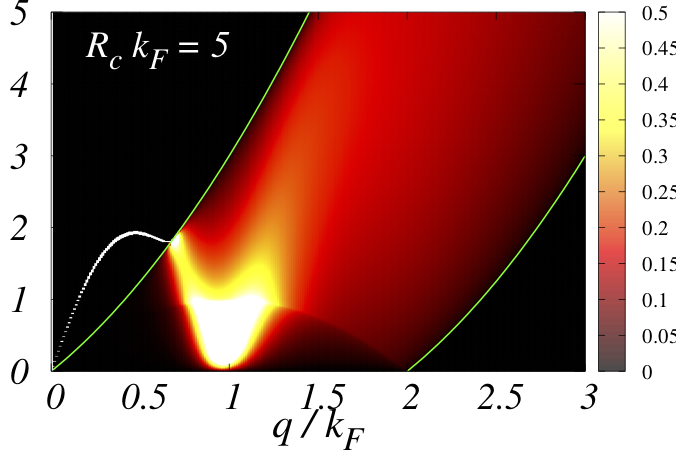}
		\end{tabular}	
	\caption{The density plots of the dynamical structure factor (in units of $\hbar/\varepsilon_{\rm F}$) versus $q/ k_{\rm F}$ and $\hbar\omega/\varepsilon_{\rm F}$ at a fixed value of the interaction strength $U=3\, \varepsilon_{\rm F}$ and for two fixed values of the soft-core radius for 3D (left) and 2D (right) Rydberg-dressed Fermi liquids. The green lines show the borders of the single-particle excitation continuum. The imaginary part of the density-density response function outside the continuum is broadened by $10^{-4}$ to make the Dirac delta peak of the collective mode visible. \label{fig:skw}} 
\end{figure}

The zero-temperature dynamical structure factor is proportional to the imaginary part of the interacting density-density response function 
\be
S(q,\omega)=-\frac{\hbar}{n\pi} \Im m \chi(q,\omega).
\ee
Using a static approximation for the effective interaction, such as the one given by Eq.\,\eqref{eq:psi_eff}, the imaginary part of the density-density response function remains non-zero only inside the single-particle excitation
continuum, where the imaginary part of the non-interacting density-density response function is non-zero and along 
the dispersion of collective density mode, where it is proportional to a Dirac delta peak~\cite{giuliani2005quantum}. 
In Fig.\,\ref{fig:skw} we illustrate the dynamical structure factor for three- and two-dimensional Rydberg-dressed Fermi liquids at a fixed interaction strength and for two different values of the soft-core radius.
The broadening of the collective mode inside the single-particle excitation continuum due to the Landau damping is evident. At larger values of the soft-core radius (bottom panels) softening of the collective mode inside the continuum is an indication of the density-wave instability, which will be discussed in detail in the next subsection.

\subsection{Density-wave instability\label{DWI}}
When the static density response function of a homogeneous system diverges, or equivalently its static dielectric function becomes zero at a specific wave vector $q_I$, the homogeneous system becomes unstable towards a density modulated phase with wavelength $\lambda_I=2\pi/q_I$.  
However, one should note that this instability corresponds to a second-order phase transition in the density channel, and relies on the presumption that no first-order phase transition, or a competing second-order phase-transition in other channels, precedes it.   

Using the mean-field Hartree-Fock approximation and the above-mentioned density instability criterion from the density response function within the RPA, a metallic quantum solid phase has been predicted for a 3D system of Rydberg-dressed fermions~\cite{li2016emergence}. The mean-field method predicted a \emph{first-order} phase transition from a homogeneous phase to the BCC crystalline structure, but interestingly the phase boundary between liquid and solid phases obtained from two techniques was in very good agreement. For a two-dimensional system of Rydberg-dressed fermions, density instability has been investigated using the RPA for the density-density response function~\cite{khasseh2017phase} too.
\begin{figure} 
\includegraphics[width=0.9\linewidth]{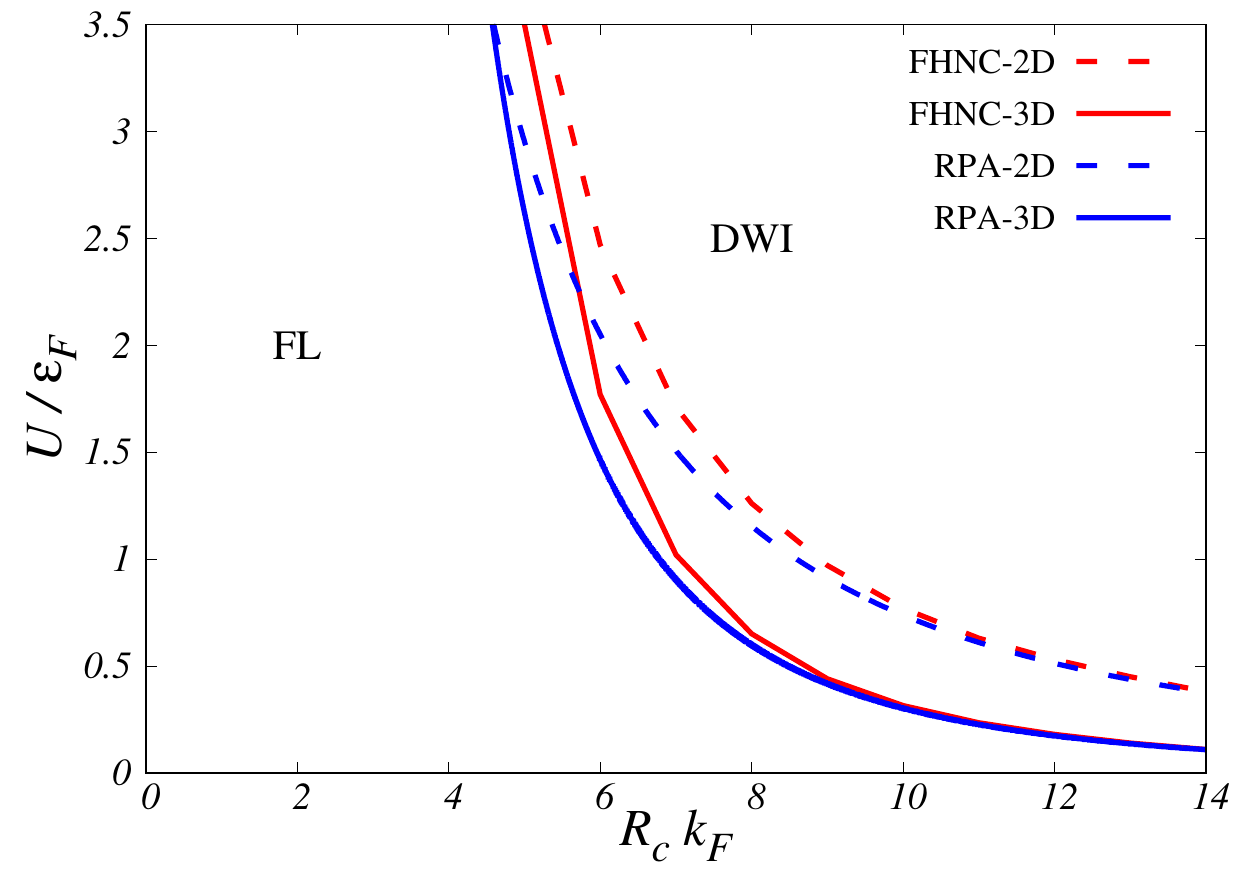} 
\caption{The phase diagram of 3D (solid lines) and 2D (dashed lines) Rydberg dressed fermions versus $u$ and $r_c$, obtained from the RPA (blue lines) and FHNC approximation (red lines). The stable homogeneous Fermi liquid  (FL) phase and regions where it becomes unstable towards the density wave instability phase (DWI) are shown in the phase diagram. 
\label{fig:PD}}
\end{figure}
In Fig.\,\ref{fig:PD}, we compare the phase diagram of Rydberg-dressed fermions obtained from the poles of the density-density response function~\eqref{eq:chi_qw} in the static limit,
within the RPA, where $V_{\rm ph}(q)$ is replaced with the bare interaction $v_{\rm RD}(q)$, and the FHNC approximation, where the effective interaction is extracted from Eq.\,\eqref{eq:psi_eff} using the numerical data for the static structure factor obtained from KP calculations.  In the later approximation, the effects of the exchange and correlation holes are approximately included in the effective interaction. 
This inclusion of exchange-correlation effects within the KP approximation makes the homogeneous liquid phase more stable. This is in line with the results obtained from the improvements over RPA with the inclusion of the Hubbard local-field factor~\cite{khasseh2017phase}. However, the phase boundary seems not to be very sensitive to the exchange-correlation at intermediate and large soft core radii. 
At small values of $r_c$, the difference between the two approximations becomes significant, and eventually, we do not find any instability at much smaller soft-core radii when we include the effects of exchange-correlation in the effective interaction.

As the static non-interacting density-density response function $\chi_0(q,\omega=0)$ is always negative, the vanishing of the denominator of Eq.\,\eqref{eq:chi_qw} is possible only in the regions of wave vector where the effective interaction is also negative. 
As the Fourier transform of the bare interaction $v_{\rm RD}(q)$ becomes negative around $q\approx 5/R_c$, the wavelength of the density modulated phase would be directly proportional to the soft-core radius within the RPA. At small enough soft-core radii, RPA unphysically predicts instability whose wavelength $\lambda_I$ is much smaller than the average distance between particles, \emph{i.e.}, $q_I >> 2 k_{\rm F}$~\cite{khasseh2017phase}. 
In contrast, when the effective interaction is extracted from the static structure factor, the instability wave vector is given by the location of the main peak in $S(q)$. As we discussed in section~\ref{sec:gr}, this is related to the average spacing between droplets in the crystalline phase, and never exceeds  $2 k_{\rm F}$. 

To obtain the lattice constant and the approximate number of atoms in each droplet, we determine the instability wave-vector from the main peak of the static structure factor and then obtain the lattice constant and the number of atoms in each droplet. We have reported the properties of representative lattice structures made of quantum droplets, 
in three- and two-dimensions in Table.~\ref{lattice}. 
In both spatial dimensions, increasing the dimensionless soft-core radius $r_c$, the lattice constant, and the number of particles in each droplet increases.

\begin{table}
\caption{Instability wave vector $q_{I}$ (in units of $k_{\rm F}$), lattice constant $a$ (in units of $1/k_{\rm F}$), and the rounded number of particles per droplet $N_d$, of Rydberg-dressed fermions in a body-centered cubic three-dimensional and a two-dimensional triangular lattice structure for several values of the soft-core radius $r_c$ and interaction strength $u$ in the vicinity of the density-wave instability.} \label{lattice}
\begin{tabular}{ |c|c|c|c|c|c| }
\hline
&$r_{c}$ & $u$&$ q_I\, [ k_{\rm F}]$&$ a\, [1 / k_{\rm F}]$&$ N_d$ \\
\hline
$3D$&$4$&  $9 $&$ 1.28$& $6.94$ & $3$ \\
\hline
$3D$&$5$& $4$&$ 1.065$& $8.34$ & $5$ \\
\hline
$3D$&$6$& $2$&$0.89$& $9.98$ & $8$\\
\hline 
$3D$&$7$& $1$&$0.77$& $11.54$ & $13$\\
\hline
$2D$&$4$& $7 $&$ 1.22$& $5.95$ & $2$ \\
\hline
$2D$&$5$& $ 4$&$0.98$& $7.4$ & $4$\\
\hline 
$2D$&$6$& $3$&$0.81$& $8.96$ & $6$\\
\hline
$2D$&$7$& $ 2$&$0.69$& $10.51$ & $8$\\
\hline
\end{tabular}
\end{table}

\section{summary}\label{sec:sum}
We have studied the ground-state properties of Rydberg-dressed Fermi liquids in two- and three-dimensions in the framework of Fermi-hypernetted-chain Euler-Lagrange approximation. The emergence of an extra small distance peak in the pair distribution function and the shift of the main peak in the static structure factor to long wavelengths with increasing interaction strength or the soft-core radius signals the phase transition from the homogenous Fermi liquid to quantum droplet crystalline phase. 
Our anticipated droplet structure regimes are experimentally accessible, especially in two-dimensions.

We have also calculated the ground-state energy of the homogeneous liquid phase. We found that the correlation energy is considerable only at intermediate values of the soft-core radii. At both small and large values of the soft-core parameter, the mean-field approximation seems to be adequate to describe the physics of the Rydberg-dressed fermions.
We should note that, to the best of our knowledge, more accurate numerical techniques such as the quantum Monte-Carlo simulations are not yet available for Rydberg-dressed fermions. However, previous experience with Fermi liquids with other forms of interactions~\cite{PhysRevB.68.155112, ASGARI2004301, ABEDINPOUR201425}, suggests that the KP results for the ground state energy are generally reliable up to very strong couplings. The relative error in comparison to the exact results is not expected to exceed a few percent.  

We should also note that we have examined other approximate methods such as the RPA and Singwi-Tosi-Land-Sj{\"o}lander (STLS) approximations~\cite{PhysRev.176.589} to find the ground state properties and the correlation energy of Rydberg-dresses Fermi liquids~\cite{Seydi_phd}. Notably, within the RPA the correlation energy is strongly overestimated. This leads to an erroneous prediction of the self-bound state for Rydberg-dressed fermions. 
The STLS approximation, although largely improves the RPA results for the ground-state energy, even at intermediate couplings its pair distribution function severely violates the exact conditions, \emph{i.e.}, the positivity and vanishing on-top value~\cite{Seydi_phd}.


\acknowledgements
We thank Robert E. Zillich, Eugene Demler, and Andy Schofield for very fruitful discussions. 
SHA is supported by Iran Science Elites Federation (ISEF). 
BT acknowledges the support from (The Scientific and Technological Research Council of Turkey) TUBITAK and (Turkish Academy of Sciences) TUBA.

\appendix
\section{Comparison between Kallio-Piilo and ladder+ approximations for the effective interaction\label{sec:panholzer}}
\begin{figure}
	\includegraphics[width=0.5\textwidth]{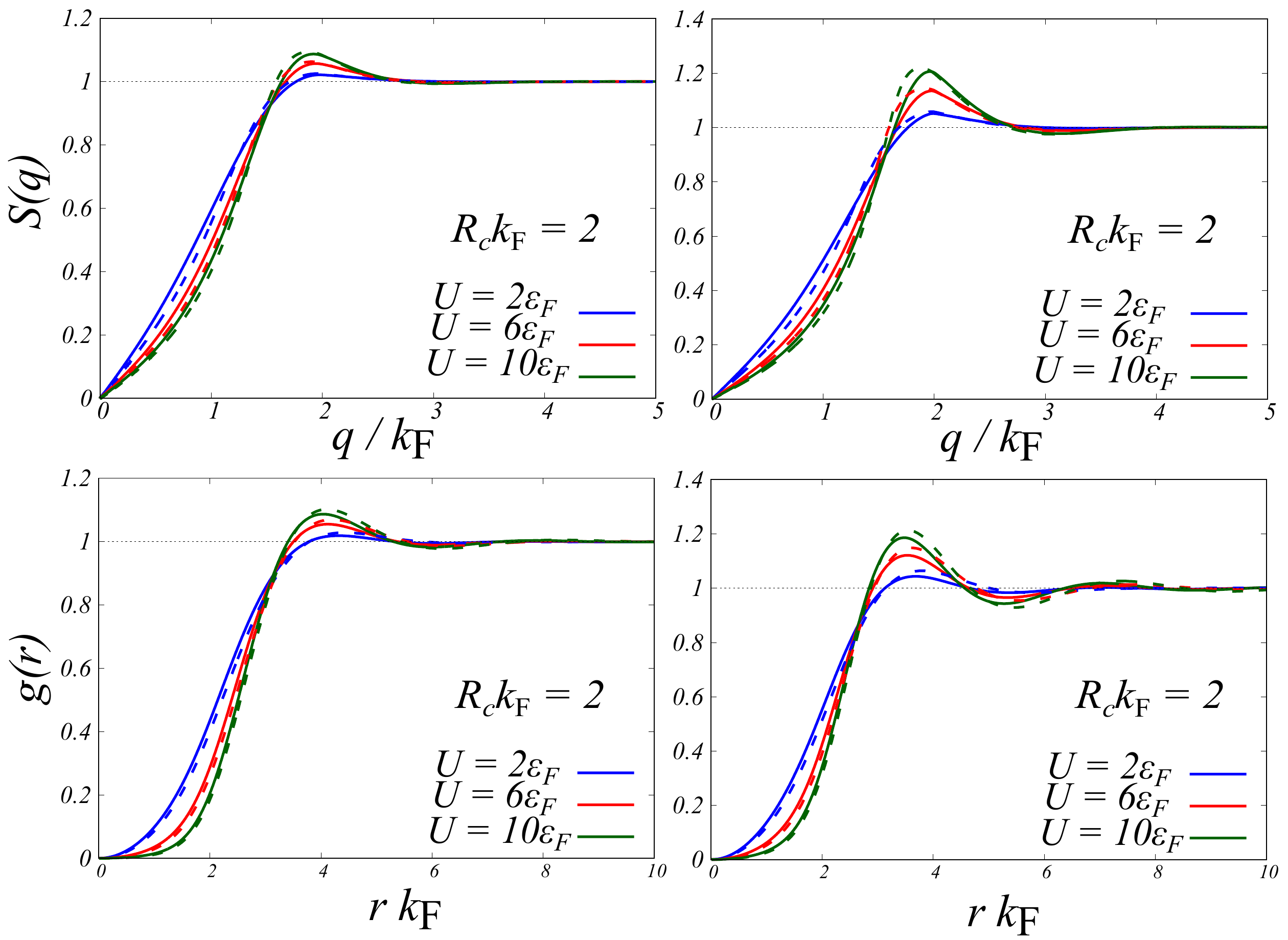}
	\caption{Comparison between the results of KP (solid lines) and PHB (dashed lines) approximations for the Fermi potentials of FHNC method, for static structure factor (top) and pair distribution function (bottom) at $r_c = 2$ and several coupling strengths for 3D (left) and 2D (right) systems of Rydberg dressed fermions. 
			\label{fig:comparison}}
\end{figure}

Recently, Panholzer, Hobbiger, and B{\"o}hm (PHB), proposed a new particle-hole effective potential for the Fermi-hypernetted chain approximation, based on the approximate summation of ladder and ring diagrams~\cite{panholzer2018optimized}. The self-consistent equations of PHB essentially become identical to the KP equations, if one replaces the Fermi potential $w_{\rm F}(r)$ in equation~\eqref{eq:W_F} with
\begin{equation}
w^{\rm PHB}_{\rm F}(r) =\frac{\hbar^2}{m}  \frac{\nabla_{\textbf{r}}^{2} \sqrt{g_{0}(r)}}{\sqrt{g_{0}(r)}}+ w_l(r),
\end{equation}
where
\be
 w_l(q)=\frac{\varepsilon_q}{2 n}\left[2 S(q)\left(1-\frac{1}{S^3_0(q)}\right)-3\left(1-\frac{1}{S^2_0(q)}\right)\right].
\ee
In Fig.\,\ref{fig:comparison}, we compare the results for the static structure factor and pair distribution function of Rydberg-dressed fermions, obtained from the KP and PHB approximations.
Both methods give very similar results for the set of parameters we have checked here. The PHB method gives slightly more pronounced peaks for both the static structure factor and the pair distribution function at strong couplings. Validation of both methods and their performance at different coupling strengths requires more accurate results \emph{e.g.}, the ones obtained from the quantum Monte Carlo simulations. Such benchmark data are not yet available in the literature. 

\bibliography{reference.bib}

\begin{thebibliography}{51}%
\makeatletter
\providecommand \@ifxundefined [1]{%
 \@ifx{#1\undefined}
}%
\providecommand \@ifnum [1]{%
 \ifnum #1\expandafter \@firstoftwo
 \else \expandafter \@secondoftwo
 \fi
}%
\providecommand \@ifx [1]{%
 \ifx #1\expandafter \@firstoftwo
 \else \expandafter \@secondoftwo
 \fi
}%
\providecommand \natexlab [1]{#1}%
\providecommand \enquote  [1]{``#1''}%
\providecommand \bibnamefont  [1]{#1}%
\providecommand \bibfnamefont [1]{#1}%
\providecommand \citenamefont [1]{#1}%
\providecommand \href@noop [0]{\@secondoftwo}%
\providecommand \href [0]{\begingroup \@sanitize@url \@href}%
\providecommand \@href[1]{\@@startlink{#1}\@@href}%
\providecommand \@@href[1]{\endgroup#1\@@endlink}%
\providecommand \@sanitize@url [0]{\catcode `\\12\catcode `\$12\catcode
  `\&12\catcode `\#12\catcode `\^12\catcode `\_12\catcode `\%12\relax}%
\providecommand \@@startlink[1]{}%
\providecommand \@@endlink[0]{}%
\providecommand \url  [0]{\begingroup\@sanitize@url \@url }%
\providecommand \@url [1]{\endgroup\@href {#1}{\urlprefix }}%
\providecommand \urlprefix  [0]{URL }%
\providecommand \Eprint [0]{\href }%
\providecommand \doibase [0]{http://dx.doi.org/}%
\providecommand \selectlanguage [0]{\@gobble}%
\providecommand \bibinfo  [0]{\@secondoftwo}%
\providecommand \bibfield  [0]{\@secondoftwo}%
\providecommand \translation [1]{[#1]}%
\providecommand \BibitemOpen [0]{}%
\providecommand \bibitemStop [0]{}%
\providecommand \bibitemNoStop [0]{.\EOS\space}%
\providecommand \EOS [0]{\spacefactor3000\relax}%
\providecommand \BibitemShut  [1]{\csname bibitem#1\endcsname}%
\let\auto@bib@innerbib\@empty
\bibitem [{\citenamefont {Bernien}\ \emph {et~al.}(2017)\citenamefont
  {Bernien}, \citenamefont {Schwartz}, \citenamefont {Keesling}, \citenamefont
  {Levine}, \citenamefont {Omran}, \citenamefont {Pichler}, \citenamefont
  {Choi}, \citenamefont {Zibrov}, \citenamefont {Endres}, \citenamefont
  {Greiner}, \citenamefont {Vuleti{\'{c}}},\ and\ \citenamefont
  {Lukin}}]{bernien2017}%
  \BibitemOpen
  \bibfield  {author} {\bibinfo {author} {\bibfnamefont {H.}~\bibnamefont
  {Bernien}}, \bibinfo {author} {\bibfnamefont {S.}~\bibnamefont {Schwartz}},
  \bibinfo {author} {\bibfnamefont {A.}~\bibnamefont {Keesling}}, \bibinfo
  {author} {\bibfnamefont {H.}~\bibnamefont {Levine}}, \bibinfo {author}
  {\bibfnamefont {A.}~\bibnamefont {Omran}}, \bibinfo {author} {\bibfnamefont
  {H.}~\bibnamefont {Pichler}}, \bibinfo {author} {\bibfnamefont
  {S.}~\bibnamefont {Choi}}, \bibinfo {author} {\bibfnamefont {A.~S.}\
  \bibnamefont {Zibrov}}, \bibinfo {author} {\bibfnamefont {M.}~\bibnamefont
  {Endres}}, \bibinfo {author} {\bibfnamefont {M.}~\bibnamefont {Greiner}},
  \bibinfo {author} {\bibfnamefont {V.}~\bibnamefont {Vuleti{\'{c}}}}, \ and\
  \bibinfo {author} {\bibfnamefont {M.~D.}\ \bibnamefont {Lukin}},\ }\href
  {\doibase 10.1038/nature24622} {\bibfield  {journal} {\bibinfo  {journal}
  {Nature}\ }\textbf {\bibinfo {volume} {551}},\ \bibinfo {pages} {579}
  (\bibinfo {year} {2017})}\BibitemShut {NoStop}%
\bibitem [{\citenamefont {Zeiher}\ \emph {et~al.}(2016)\citenamefont {Zeiher},
  \citenamefont {Van~Bijnen}, \citenamefont {Schau{\ss}}, \citenamefont {Hild},
  \citenamefont {Choi}, \citenamefont {Pohl}, \citenamefont {Bloch},\ and\
  \citenamefont {Gross}}]{zeiher2016many}%
  \BibitemOpen
  \bibfield  {author} {\bibinfo {author} {\bibfnamefont {J.}~\bibnamefont
  {Zeiher}}, \bibinfo {author} {\bibfnamefont {R.}~\bibnamefont {Van~Bijnen}},
  \bibinfo {author} {\bibfnamefont {P.}~\bibnamefont {Schau{\ss}}}, \bibinfo
  {author} {\bibfnamefont {S.}~\bibnamefont {Hild}}, \bibinfo {author}
  {\bibfnamefont {J.-y.}\ \bibnamefont {Choi}}, \bibinfo {author}
  {\bibfnamefont {T.}~\bibnamefont {Pohl}}, \bibinfo {author} {\bibfnamefont
  {I.}~\bibnamefont {Bloch}}, \ and\ \bibinfo {author} {\bibfnamefont
  {C.}~\bibnamefont {Gross}},\ }\href {\doibase 10.1038/nphys3835} {\bibfield
  {journal} {\bibinfo  {journal} {Nature Phys.}\ }\textbf {\bibinfo {volume}
  {12}},\ \bibinfo {pages} {1095} (\bibinfo {year} {2016})}\BibitemShut
  {NoStop}%
\bibitem [{\citenamefont {Labuhn}\ \emph {et~al.}(2016)\citenamefont {Labuhn},
  \citenamefont {Barredo}, \citenamefont {Ravets}, \citenamefont
  {de~L{\'{e}}s{\'{e}}leuc}, \citenamefont {Macr{\`{\i}}}, \citenamefont
  {Lahaye},\ and\ \citenamefont {Browaeys}}]{labuhn2016tunable}%
  \BibitemOpen
  \bibfield  {author} {\bibinfo {author} {\bibfnamefont {H.}~\bibnamefont
  {Labuhn}}, \bibinfo {author} {\bibfnamefont {D.}~\bibnamefont {Barredo}},
  \bibinfo {author} {\bibfnamefont {S.}~\bibnamefont {Ravets}}, \bibinfo
  {author} {\bibfnamefont {S.}~\bibnamefont {de~L{\'{e}}s{\'{e}}leuc}},
  \bibinfo {author} {\bibfnamefont {T.}~\bibnamefont {Macr{\`{\i}}}}, \bibinfo
  {author} {\bibfnamefont {T.}~\bibnamefont {Lahaye}}, \ and\ \bibinfo {author}
  {\bibfnamefont {A.}~\bibnamefont {Browaeys}},\ }\href {\doibase
  10.1038/nature18274} {\bibfield  {journal} {\bibinfo  {journal} {Nature}\
  }\textbf {\bibinfo {volume} {534}},\ \bibinfo {pages} {667} (\bibinfo {year}
  {2016})}\BibitemShut {NoStop}%
\bibitem [{\citenamefont {Balewski}\ \emph {et~al.}(2014)\citenamefont
  {Balewski}, \citenamefont {Krupp}, \citenamefont {Gaj}, \citenamefont
  {Hofferberth}, \citenamefont {L{\"o}w},\ and\ \citenamefont
  {Pfau}}]{balewski2014NewJournalofPhysics}%
  \BibitemOpen
  \bibfield  {author} {\bibinfo {author} {\bibfnamefont {J.~B.}\ \bibnamefont
  {Balewski}}, \bibinfo {author} {\bibfnamefont {A.~T.}\ \bibnamefont {Krupp}},
  \bibinfo {author} {\bibfnamefont {A.}~\bibnamefont {Gaj}}, \bibinfo {author}
  {\bibfnamefont {S.}~\bibnamefont {Hofferberth}}, \bibinfo {author}
  {\bibfnamefont {R.}~\bibnamefont {L{\"o}w}}, \ and\ \bibinfo {author}
  {\bibfnamefont {T.}~\bibnamefont {Pfau}},\ }\href {\doibase
  10.1088/1367-2630/16/6/063012} {\bibfield  {journal} {\bibinfo  {journal}
  {New J. Phys.}\ }\textbf {\bibinfo {volume} {16}},\ \bibinfo {pages} {063012}
  (\bibinfo {year} {2014})}\BibitemShut {NoStop}%
\bibitem [{\citenamefont {Firstenberg}\ \emph {et~al.}(2016)\citenamefont
  {Firstenberg}, \citenamefont {Adams},\ and\ \citenamefont
  {Hofferberth}}]{firstenberg2016nonlinear}%
  \BibitemOpen
  \bibfield  {author} {\bibinfo {author} {\bibfnamefont {O.}~\bibnamefont
  {Firstenberg}}, \bibinfo {author} {\bibfnamefont {C.~S.}\ \bibnamefont
  {Adams}}, \ and\ \bibinfo {author} {\bibfnamefont {S.}~\bibnamefont
  {Hofferberth}},\ }\href {\doibase 10.1088/0953-4075/49/15/152003} {\bibfield
  {journal} {\bibinfo  {journal} {J. Phys. B: At. Mol. Opt. Phys.}\ }\textbf
  {\bibinfo {volume} {49}},\ \bibinfo {pages} {152003} (\bibinfo {year}
  {2016})}\BibitemShut {NoStop}%
\bibitem [{\citenamefont {Dudin}\ and\ \citenamefont
  {Kuzmich}(2012)}]{dudin2012strongly}%
  \BibitemOpen
  \bibfield  {author} {\bibinfo {author} {\bibfnamefont {Y.}~\bibnamefont
  {Dudin}}\ and\ \bibinfo {author} {\bibfnamefont {A.}~\bibnamefont
  {Kuzmich}},\ }\href {https://science.sciencemag.org/content/336/6083/887}
  {\bibfield  {journal} {\bibinfo  {journal} {Science}\ }\textbf {\bibinfo
  {volume} {336}},\ \bibinfo {pages} {887} (\bibinfo {year}
  {2012})}\BibitemShut {NoStop}%
\bibitem [{\citenamefont {Adams}\ \emph {et~al.}(2019)\citenamefont {Adams},
  \citenamefont {Pritchard},\ and\ \citenamefont {Shaffer}}]{Adams_2019}%
  \BibitemOpen
  \bibfield  {author} {\bibinfo {author} {\bibfnamefont {C.~S.}\ \bibnamefont
  {Adams}}, \bibinfo {author} {\bibfnamefont {J.~D.}\ \bibnamefont
  {Pritchard}}, \ and\ \bibinfo {author} {\bibfnamefont {J.~P.}\ \bibnamefont
  {Shaffer}},\ }\href {\doibase 10.1088/1361-6455/ab52ef} {\bibfield  {journal}
  {\bibinfo  {journal} {J. Phys. B: At. Mol. Opt. Phys.}\ }\textbf {\bibinfo
  {volume} {53}},\ \bibinfo {pages} {012002} (\bibinfo {year}
  {2019})}\BibitemShut {NoStop}%
\bibitem [{\citenamefont {Saffman}\ \emph {et~al.}(2010)\citenamefont
  {Saffman}, \citenamefont {Walker},\ and\ \citenamefont
  {M\o{}lmer}}]{saffman2010quantum}%
  \BibitemOpen
  \bibfield  {author} {\bibinfo {author} {\bibfnamefont {M.}~\bibnamefont
  {Saffman}}, \bibinfo {author} {\bibfnamefont {T.~G.}\ \bibnamefont {Walker}},
  \ and\ \bibinfo {author} {\bibfnamefont {K.}~\bibnamefont {M\o{}lmer}},\
  }\href {\doibase 10.1103/RevModPhys.82.2313} {\bibfield  {journal} {\bibinfo
  {journal} {Rev. Mod. Phys.}\ }\textbf {\bibinfo {volume} {82}},\ \bibinfo
  {pages} {2313} (\bibinfo {year} {2010})}\BibitemShut {NoStop}%
\bibitem [{\citenamefont {Isenhower}\ \emph {et~al.}(2010)\citenamefont
  {Isenhower}, \citenamefont {Urban}, \citenamefont {Zhang}, \citenamefont
  {Gill}, \citenamefont {Henage}, \citenamefont {Johnson}, \citenamefont
  {Walker},\ and\ \citenamefont {Saffman}}]{isenhower2010demonstration}%
  \BibitemOpen
  \bibfield  {author} {\bibinfo {author} {\bibfnamefont {L.}~\bibnamefont
  {Isenhower}}, \bibinfo {author} {\bibfnamefont {E.}~\bibnamefont {Urban}},
  \bibinfo {author} {\bibfnamefont {X.~L.}\ \bibnamefont {Zhang}}, \bibinfo
  {author} {\bibfnamefont {A.~T.}\ \bibnamefont {Gill}}, \bibinfo {author}
  {\bibfnamefont {T.}~\bibnamefont {Henage}}, \bibinfo {author} {\bibfnamefont
  {T.~A.}\ \bibnamefont {Johnson}}, \bibinfo {author} {\bibfnamefont {T.~G.}\
  \bibnamefont {Walker}}, \ and\ \bibinfo {author} {\bibfnamefont
  {M.}~\bibnamefont {Saffman}},\ }\href {\doibase
  10.1103/PhysRevLett.104.010503} {\bibfield  {journal} {\bibinfo  {journal}
  {Phys. Rev. Lett.}\ }\textbf {\bibinfo {volume} {104}},\ \bibinfo {pages}
  {010503} (\bibinfo {year} {2010})}\BibitemShut {NoStop}%
\bibitem [{\citenamefont {Lukin}\ \emph {et~al.}(2001)\citenamefont {Lukin},
  \citenamefont {Fleischhauer}, \citenamefont {Cote}, \citenamefont {Duan},
  \citenamefont {Jaksch}, \citenamefont {Cirac},\ and\ \citenamefont
  {Zoller}}]{lukin2001dipole}%
  \BibitemOpen
  \bibfield  {author} {\bibinfo {author} {\bibfnamefont {M.~D.}\ \bibnamefont
  {Lukin}}, \bibinfo {author} {\bibfnamefont {M.}~\bibnamefont {Fleischhauer}},
  \bibinfo {author} {\bibfnamefont {R.}~\bibnamefont {Cote}}, \bibinfo {author}
  {\bibfnamefont {L.~M.}\ \bibnamefont {Duan}}, \bibinfo {author}
  {\bibfnamefont {D.}~\bibnamefont {Jaksch}}, \bibinfo {author} {\bibfnamefont
  {J.~I.}\ \bibnamefont {Cirac}}, \ and\ \bibinfo {author} {\bibfnamefont
  {P.}~\bibnamefont {Zoller}},\ }\href
  {https://journals.aps.org/prl/abstract/10.1103/PhysRevLett.87.037901}
  {\bibfield  {journal} {\bibinfo  {journal} {Phys. Rev. Lett.}\ }\textbf
  {\bibinfo {volume} {87}},\ \bibinfo {pages} {037901} (\bibinfo {year}
  {2001})}\BibitemShut {NoStop}%
\bibitem [{\citenamefont {Browaeys}\ and\ \citenamefont
  {Lahaye}(2016)}]{browaeys2016interacting}%
  \BibitemOpen
  \bibfield  {author} {\bibinfo {author} {\bibfnamefont {A.}~\bibnamefont
  {Browaeys}}\ and\ \bibinfo {author} {\bibfnamefont {T.}~\bibnamefont
  {Lahaye}},\ }in\ \href
  {https://link.springer.com/chapter/10.1007%2F978-3-319-14316-3_7} {\emph
  {\bibinfo {booktitle} {Niels Bohr, 1913-2013}}}\ (\bibinfo  {publisher}
  {Springer},\ \bibinfo {year} {2016})\ pp.\ \bibinfo {pages}
  {177--198}\BibitemShut {NoStop}%
\bibitem [{\citenamefont {Weimer}\ \emph {et~al.}(2010)\citenamefont {Weimer},
  \citenamefont {M{\"u}ller}, \citenamefont {Lesanovsky}, \citenamefont
  {Zoller},\ and\ \citenamefont {B{\"u}chler}}]{weimer2010rydberg}%
  \BibitemOpen
  \bibfield  {author} {\bibinfo {author} {\bibfnamefont {H.}~\bibnamefont
  {Weimer}}, \bibinfo {author} {\bibfnamefont {M.}~\bibnamefont {M{\"u}ller}},
  \bibinfo {author} {\bibfnamefont {I.}~\bibnamefont {Lesanovsky}}, \bibinfo
  {author} {\bibfnamefont {P.}~\bibnamefont {Zoller}}, \ and\ \bibinfo {author}
  {\bibfnamefont {H.~P.}\ \bibnamefont {B{\"u}chler}},\ }\href {\doibase
  10.1038/nphys1614} {\bibfield  {journal} {\bibinfo  {journal} {Nature Phys.}\
  }\textbf {\bibinfo {volume} {6}},\ \bibinfo {pages} {382} (\bibinfo {year}
  {2010})}\BibitemShut {NoStop}%
\bibitem [{\citenamefont {P{\l}odzie{\'n}}\ \emph {et~al.}(2018)\citenamefont
  {P{\l}odzie{\'n}}, \citenamefont {Sowi{\'n}ski},\ and\ \citenamefont
  {Kokkelmans}}]{plodzien2018scientificreports}%
  \BibitemOpen
  \bibfield  {author} {\bibinfo {author} {\bibfnamefont {M.}~\bibnamefont
  {P{\l}odzie{\'n}}}, \bibinfo {author} {\bibfnamefont {T.}~\bibnamefont
  {Sowi{\'n}ski}}, \ and\ \bibinfo {author} {\bibfnamefont {S.}~\bibnamefont
  {Kokkelmans}},\ }\href {https://www.nature.com/articles/s41598-018-27232-4}
  {\bibfield  {journal} {\bibinfo  {journal} {Scientific reports}\ }\textbf
  {\bibinfo {volume} {8}},\ \bibinfo {pages} {9247} (\bibinfo {year}
  {2018})}\BibitemShut {NoStop}%
\bibitem [{\citenamefont {Karpiuk}\ \emph {et~al.}(2015)\citenamefont
  {Karpiuk}, \citenamefont {Brewczyk}, \citenamefont {Rz{{a}}{\.{z}}ewski},
  \citenamefont {Gaj}, \citenamefont {Balewski}, \citenamefont {Krupp},
  \citenamefont {Schlagmüller}, \citenamefont {Löw}, \citenamefont
  {Hofferberth},\ and\ \citenamefont {Pfau}}]{karpiuk2015imaging}%
  \BibitemOpen
  \bibfield  {author} {\bibinfo {author} {\bibfnamefont {T.}~\bibnamefont
  {Karpiuk}}, \bibinfo {author} {\bibfnamefont {M.}~\bibnamefont {Brewczyk}},
  \bibinfo {author} {\bibfnamefont {K.}~\bibnamefont {Rz{{a}}{\.{z}}ewski}},
  \bibinfo {author} {\bibfnamefont {A.}~\bibnamefont {Gaj}}, \bibinfo {author}
  {\bibfnamefont {J.~B.}\ \bibnamefont {Balewski}}, \bibinfo {author}
  {\bibfnamefont {A.~T.}\ \bibnamefont {Krupp}}, \bibinfo {author}
  {\bibfnamefont {M.}~\bibnamefont {Schlagmüller}}, \bibinfo {author}
  {\bibfnamefont {R.}~\bibnamefont {Löw}}, \bibinfo {author} {\bibfnamefont
  {S.}~\bibnamefont {Hofferberth}}, \ and\ \bibinfo {author} {\bibfnamefont
  {T.}~\bibnamefont {Pfau}},\ }\href {\doibase 10.1088/1367-2630/17/5/053046}
  {\bibfield  {journal} {\bibinfo  {journal} {New J. Phys.}\ }\textbf {\bibinfo
  {volume} {17}},\ \bibinfo {pages} {053046} (\bibinfo {year}
  {2015})}\BibitemShut {NoStop}%
\bibitem [{\citenamefont {Henkel}\ \emph {et~al.}(2010)\citenamefont {Henkel},
  \citenamefont {Nath},\ and\ \citenamefont {Pohl}}]{henkel2010three}%
  \BibitemOpen
  \bibfield  {author} {\bibinfo {author} {\bibfnamefont {N.}~\bibnamefont
  {Henkel}}, \bibinfo {author} {\bibfnamefont {R.}~\bibnamefont {Nath}}, \ and\
  \bibinfo {author} {\bibfnamefont {T.}~\bibnamefont {Pohl}},\ }\href {\doibase
  10.1103/PhysRevLett.104.195302} {\bibfield  {journal} {\bibinfo  {journal}
  {Phys. Rev. Lett.}\ }\textbf {\bibinfo {volume} {104}},\ \bibinfo {pages}
  {195302} (\bibinfo {year} {2010})}\BibitemShut {NoStop}%
\bibitem [{\citenamefont {Browaeys}\ \emph {et~al.}(2016)\citenamefont
  {Browaeys}, \citenamefont {Barredo},\ and\ \citenamefont
  {Lahaye}}]{browaeys2016experimental}%
  \BibitemOpen
  \bibfield  {author} {\bibinfo {author} {\bibfnamefont {A.}~\bibnamefont
  {Browaeys}}, \bibinfo {author} {\bibfnamefont {D.}~\bibnamefont {Barredo}}, \
  and\ \bibinfo {author} {\bibfnamefont {T.}~\bibnamefont {Lahaye}},\ }\href
  {http://stacks.iop.org/0953-4075/49/i=15/a=152001} {\bibfield  {journal}
  {\bibinfo  {journal} {J. Phys. B: At. Mol. Opt. Phys.}\ }\textbf {\bibinfo
  {volume} {49}},\ \bibinfo {pages} {152001} (\bibinfo {year}
  {2016})}\BibitemShut {NoStop}%
\bibitem [{\citenamefont {Jau}\ \emph {et~al.}(2016)\citenamefont {Jau},
  \citenamefont {Hankin}, \citenamefont {Keating}, \citenamefont {Deutsch},\
  and\ \citenamefont {Biedermann}}]{jau2016entangling}%
  \BibitemOpen
  \bibfield  {author} {\bibinfo {author} {\bibfnamefont {Y.-Y.}\ \bibnamefont
  {Jau}}, \bibinfo {author} {\bibfnamefont {A.}~\bibnamefont {Hankin}},
  \bibinfo {author} {\bibfnamefont {T.}~\bibnamefont {Keating}}, \bibinfo
  {author} {\bibfnamefont {I.}~\bibnamefont {Deutsch}}, \ and\ \bibinfo
  {author} {\bibfnamefont {G.}~\bibnamefont {Biedermann}},\ }\href
  {https://www.nature.com/articles/nphys3487} {\bibfield  {journal} {\bibinfo
  {journal} {Nature Phys.}\ }\textbf {\bibinfo {volume} {12}},\ \bibinfo
  {pages} {71} (\bibinfo {year} {2016})}\BibitemShut {NoStop}%
\bibitem [{\citenamefont {Zeiher}\ \emph {et~al.}(2017)\citenamefont {Zeiher},
  \citenamefont {Choi}, \citenamefont {Rubio-Abadal}, \citenamefont {Pohl},
  \citenamefont {van Bijnen}, \citenamefont {Bloch},\ and\ \citenamefont
  {Gross}}]{zeiher2017coherent}%
  \BibitemOpen
  \bibfield  {author} {\bibinfo {author} {\bibfnamefont {J.}~\bibnamefont
  {Zeiher}}, \bibinfo {author} {\bibfnamefont {J.-y.}\ \bibnamefont {Choi}},
  \bibinfo {author} {\bibfnamefont {A.}~\bibnamefont {Rubio-Abadal}}, \bibinfo
  {author} {\bibfnamefont {T.}~\bibnamefont {Pohl}}, \bibinfo {author}
  {\bibfnamefont {R.}~\bibnamefont {van Bijnen}}, \bibinfo {author}
  {\bibfnamefont {I.}~\bibnamefont {Bloch}}, \ and\ \bibinfo {author}
  {\bibfnamefont {C.}~\bibnamefont {Gross}},\ }\href {\doibase
  10.1103/PhysRevX.7.041063} {\bibfield  {journal} {\bibinfo  {journal} {Phys.
  Rev. X}\ }\textbf {\bibinfo {volume} {7}},\ \bibinfo {pages} {041063}
  (\bibinfo {year} {2017})}\BibitemShut {NoStop}%
\bibitem [{\citenamefont {Hollerith}\ \emph {et~al.}(2019)\citenamefont
  {Hollerith}, \citenamefont {Zeiher}, \citenamefont {Rui}, \citenamefont
  {Rubio-Abadal}, \citenamefont {Walther}, \citenamefont {Pohl}, \citenamefont
  {Stamper-Kurn}, \citenamefont {Bloch},\ and\ \citenamefont
  {Gross}}]{Hollerith2019}%
  \BibitemOpen
  \bibfield  {author} {\bibinfo {author} {\bibfnamefont {S.}~\bibnamefont
  {Hollerith}}, \bibinfo {author} {\bibfnamefont {J.}~\bibnamefont {Zeiher}},
  \bibinfo {author} {\bibfnamefont {J.}~\bibnamefont {Rui}}, \bibinfo {author}
  {\bibfnamefont {A.}~\bibnamefont {Rubio-Abadal}}, \bibinfo {author}
  {\bibfnamefont {V.}~\bibnamefont {Walther}}, \bibinfo {author} {\bibfnamefont
  {T.}~\bibnamefont {Pohl}}, \bibinfo {author} {\bibfnamefont {D.~M.}\
  \bibnamefont {Stamper-Kurn}}, \bibinfo {author} {\bibfnamefont
  {I.}~\bibnamefont {Bloch}}, \ and\ \bibinfo {author} {\bibfnamefont
  {C.}~\bibnamefont {Gross}},\ }\href
  {https://science.sciencemag.org/content/364/6441/664.abstract} {\bibfield
  {journal} {\bibinfo  {journal} {Science}\ }\textbf {\bibinfo {volume}
  {364}},\ \bibinfo {pages} {664} (\bibinfo {year} {2019})}\BibitemShut
  {NoStop}%
\bibitem [{\citenamefont {Pupillo}\ \emph {et~al.}(2010)\citenamefont
  {Pupillo}, \citenamefont {Micheli}, \citenamefont {Boninsegni}, \citenamefont
  {Lesanovsky},\ and\ \citenamefont {Zoller}}]{pupillo2010strongly}%
  \BibitemOpen
  \bibfield  {author} {\bibinfo {author} {\bibfnamefont {G.}~\bibnamefont
  {Pupillo}}, \bibinfo {author} {\bibfnamefont {A.}~\bibnamefont {Micheli}},
  \bibinfo {author} {\bibfnamefont {M.}~\bibnamefont {Boninsegni}}, \bibinfo
  {author} {\bibfnamefont {I.}~\bibnamefont {Lesanovsky}}, \ and\ \bibinfo
  {author} {\bibfnamefont {P.}~\bibnamefont {Zoller}},\ }\href {\doibase
  10.1103/PhysRevLett.104.223002} {\bibfield  {journal} {\bibinfo  {journal}
  {Phys. Rev. Lett.}\ }\textbf {\bibinfo {volume} {104}},\ \bibinfo {pages}
  {223002} (\bibinfo {year} {2010})}\BibitemShut {NoStop}%
\bibitem [{\citenamefont {Boninsegni}\ and\ \citenamefont
  {Prokof'ev}(2012)}]{boninsegni2012colloquium}%
  \BibitemOpen
  \bibfield  {author} {\bibinfo {author} {\bibfnamefont {M.}~\bibnamefont
  {Boninsegni}}\ and\ \bibinfo {author} {\bibfnamefont {N.~V.}\ \bibnamefont
  {Prokof'ev}},\ }\href {\doibase 10.1103/RevModPhys.84.759} {\bibfield
  {journal} {\bibinfo  {journal} {Rev. Mod. Phys.}\ }\textbf {\bibinfo {volume}
  {84}},\ \bibinfo {pages} {759} (\bibinfo {year} {2012})}\BibitemShut
  {NoStop}%
\bibitem [{\citenamefont {Cinti}\ \emph {et~al.}(2010)\citenamefont {Cinti},
  \citenamefont {Jain}, \citenamefont {Boninsegni}, \citenamefont {Micheli},
  \citenamefont {Zoller},\ and\ \citenamefont
  {Pupillo}}]{PhysRevLett.105.135301}%
  \BibitemOpen
  \bibfield  {author} {\bibinfo {author} {\bibfnamefont {F.}~\bibnamefont
  {Cinti}}, \bibinfo {author} {\bibfnamefont {P.}~\bibnamefont {Jain}},
  \bibinfo {author} {\bibfnamefont {M.}~\bibnamefont {Boninsegni}}, \bibinfo
  {author} {\bibfnamefont {A.}~\bibnamefont {Micheli}}, \bibinfo {author}
  {\bibfnamefont {P.}~\bibnamefont {Zoller}}, \ and\ \bibinfo {author}
  {\bibfnamefont {G.}~\bibnamefont {Pupillo}},\ }\href {\doibase
  10.1103/PhysRevLett.105.135301} {\bibfield  {journal} {\bibinfo  {journal}
  {Phys. Rev. Lett.}\ }\textbf {\bibinfo {volume} {105}},\ \bibinfo {pages}
  {135301} (\bibinfo {year} {2010})}\BibitemShut {NoStop}%
\bibitem [{\citenamefont {Cinti}\ \emph {et~al.}(2014)\citenamefont {Cinti},
  \citenamefont {Macrì}, \citenamefont {Lechner}, \citenamefont {Pupillo},\
  and\ \citenamefont {Pohl}}]{Cinti}%
  \BibitemOpen
  \bibfield  {author} {\bibinfo {author} {\bibfnamefont {F.}~\bibnamefont
  {Cinti}}, \bibinfo {author} {\bibfnamefont {T.}~\bibnamefont {Macrì}},
  \bibinfo {author} {\bibfnamefont {W.}~\bibnamefont {Lechner}}, \bibinfo
  {author} {\bibfnamefont {G.}~\bibnamefont {Pupillo}}, \ and\ \bibinfo
  {author} {\bibfnamefont {T.}~\bibnamefont {Pohl}},\ }\href
  {https://doi.org/10.1038/ncomms4235} {\bibfield  {journal} {\bibinfo
  {journal} {Nat. Commun..}\ }\textbf {\bibinfo {volume} {5}} (\bibinfo {year}
  {2014})}\BibitemShut {NoStop}%
\bibitem [{\citenamefont {Prestipino}\ \emph {et~al.}(2018)\citenamefont
  {Prestipino}, \citenamefont {Sergi},\ and\ \citenamefont
  {Bruno}}]{prestipino2018prb}%
  \BibitemOpen
  \bibfield  {author} {\bibinfo {author} {\bibfnamefont {S.}~\bibnamefont
  {Prestipino}}, \bibinfo {author} {\bibfnamefont {A.}~\bibnamefont {Sergi}}, \
  and\ \bibinfo {author} {\bibfnamefont {E.}~\bibnamefont {Bruno}},\ }\href
  {https://journals.aps.org/prb/abstract/10.1103/PhysRevB.98.104104} {\bibfield
   {journal} {\bibinfo  {journal} {Phys. Rev. B}\ }\textbf {\bibinfo {volume}
  {98}},\ \bibinfo {pages} {104104} (\bibinfo {year} {2018})}\BibitemShut
  {NoStop}%
\bibitem [{\citenamefont {Semeghini}\ \emph {et~al.}(2018)\citenamefont
  {Semeghini}, \citenamefont {Ferioli}, \citenamefont {Masi}, \citenamefont
  {Mazzinghi}, \citenamefont {Wolswijk}, \citenamefont {Minardi}, \citenamefont
  {Modugno}, \citenamefont {Modugno}, \citenamefont {Inguscio},\ and\
  \citenamefont {Fattori}}]{PhysRevLett.120.235301}%
  \BibitemOpen
  \bibfield  {author} {\bibinfo {author} {\bibfnamefont {G.}~\bibnamefont
  {Semeghini}}, \bibinfo {author} {\bibfnamefont {G.}~\bibnamefont {Ferioli}},
  \bibinfo {author} {\bibfnamefont {L.}~\bibnamefont {Masi}}, \bibinfo {author}
  {\bibfnamefont {C.}~\bibnamefont {Mazzinghi}}, \bibinfo {author}
  {\bibfnamefont {L.}~\bibnamefont {Wolswijk}}, \bibinfo {author}
  {\bibfnamefont {F.}~\bibnamefont {Minardi}}, \bibinfo {author} {\bibfnamefont
  {M.}~\bibnamefont {Modugno}}, \bibinfo {author} {\bibfnamefont
  {G.}~\bibnamefont {Modugno}}, \bibinfo {author} {\bibfnamefont
  {M.}~\bibnamefont {Inguscio}}, \ and\ \bibinfo {author} {\bibfnamefont
  {M.}~\bibnamefont {Fattori}},\ }\href {\doibase
  10.1103/PhysRevLett.120.235301} {\bibfield  {journal} {\bibinfo  {journal}
  {Phys. Rev. Lett.}\ }\textbf {\bibinfo {volume} {120}},\ \bibinfo {pages}
  {235301} (\bibinfo {year} {2018})}\BibitemShut {NoStop}%
\bibitem [{\citenamefont {Ferrier-Barbut}\ \emph {et~al.}(2016)\citenamefont
  {Ferrier-Barbut}, \citenamefont {Kadau}, \citenamefont {Schmitt},
  \citenamefont {Wenzel},\ and\ \citenamefont {Pfau}}]{PhysRevLett.116.215301}%
  \BibitemOpen
  \bibfield  {author} {\bibinfo {author} {\bibfnamefont {I.}~\bibnamefont
  {Ferrier-Barbut}}, \bibinfo {author} {\bibfnamefont {H.}~\bibnamefont
  {Kadau}}, \bibinfo {author} {\bibfnamefont {M.}~\bibnamefont {Schmitt}},
  \bibinfo {author} {\bibfnamefont {M.}~\bibnamefont {Wenzel}}, \ and\ \bibinfo
  {author} {\bibfnamefont {T.}~\bibnamefont {Pfau}},\ }\href {\doibase
  10.1103/PhysRevLett.116.215301} {\bibfield  {journal} {\bibinfo  {journal}
  {Phys. Rev. Lett.}\ }\textbf {\bibinfo {volume} {116}},\ \bibinfo {pages}
  {215301} (\bibinfo {year} {2016})}\BibitemShut {NoStop}%
\bibitem [{\citenamefont {Cabrera}\ \emph {et~al.}(2018)\citenamefont
  {Cabrera}, \citenamefont {Tanzi}, \citenamefont {Sanz}, \citenamefont
  {Naylor}, \citenamefont {Thomas}, \citenamefont {Cheiney},\ and\
  \citenamefont {Tarruell}}]{cabrera2018quantum}%
  \BibitemOpen
  \bibfield  {author} {\bibinfo {author} {\bibfnamefont {C.}~\bibnamefont
  {Cabrera}}, \bibinfo {author} {\bibfnamefont {L.}~\bibnamefont {Tanzi}},
  \bibinfo {author} {\bibfnamefont {J.}~\bibnamefont {Sanz}}, \bibinfo {author}
  {\bibfnamefont {B.}~\bibnamefont {Naylor}}, \bibinfo {author} {\bibfnamefont
  {P.}~\bibnamefont {Thomas}}, \bibinfo {author} {\bibfnamefont
  {P.}~\bibnamefont {Cheiney}}, \ and\ \bibinfo {author} {\bibfnamefont
  {L.}~\bibnamefont {Tarruell}},\ }\href
  {https://science.sciencemag.org/content/359/6373/301} {\bibfield  {journal}
  {\bibinfo  {journal} {Science}\ }\textbf {\bibinfo {volume} {359}},\ \bibinfo
  {pages} {301} (\bibinfo {year} {2018})}\BibitemShut {NoStop}%
\bibitem [{\citenamefont {Seydi}\ \emph {et~al.}(2020)\citenamefont {Seydi},
  \citenamefont {Abedinpour}, \citenamefont {Zillich}, \citenamefont {Asgari},\
  and\ \citenamefont {Tanatar}}]{Seydi2019arxiv}%
  \BibitemOpen
  \bibfield  {author} {\bibinfo {author} {\bibfnamefont {I.}~\bibnamefont
  {Seydi}}, \bibinfo {author} {\bibfnamefont {S.~H.}\ \bibnamefont
  {Abedinpour}}, \bibinfo {author} {\bibfnamefont {R.~E.}\ \bibnamefont
  {Zillich}}, \bibinfo {author} {\bibfnamefont {R.}~\bibnamefont {Asgari}}, \
  and\ \bibinfo {author} {\bibfnamefont {B.}~\bibnamefont {Tanatar}},\ }\href
  {\doibase 10.1103/PhysRevA.101.013628} {\bibfield  {journal} {\bibinfo
  {journal} {Phys. Rev. A}\ }\textbf {\bibinfo {volume} {101}},\ \bibinfo
  {pages} {013628} (\bibinfo {year} {2020})}\BibitemShut {NoStop}%
\bibitem [{\citenamefont {Lee}\ \emph {et~al.}(2013)\citenamefont {Lee},
  \citenamefont {Gopalakrishnan},\ and\ \citenamefont
  {Lukin}}]{PhysRevLett.110.257204}%
  \BibitemOpen
  \bibfield  {author} {\bibinfo {author} {\bibfnamefont {T.~E.}\ \bibnamefont
  {Lee}}, \bibinfo {author} {\bibfnamefont {S.}~\bibnamefont {Gopalakrishnan}},
  \ and\ \bibinfo {author} {\bibfnamefont {M.~D.}\ \bibnamefont {Lukin}},\
  }\href {\doibase 10.1103/PhysRevLett.110.257204} {\bibfield  {journal}
  {\bibinfo  {journal} {Phys. Rev. Lett.}\ }\textbf {\bibinfo {volume} {110}},\
  \bibinfo {pages} {257204} (\bibinfo {year} {2013})}\BibitemShut {NoStop}%
\bibitem [{\citenamefont {Xiong}\ \emph {et~al.}(2014)\citenamefont {Xiong},
  \citenamefont {Jen},\ and\ \citenamefont {Wang}}]{xiong2014topological}%
  \BibitemOpen
  \bibfield  {author} {\bibinfo {author} {\bibfnamefont {B.}~\bibnamefont
  {Xiong}}, \bibinfo {author} {\bibfnamefont {H.~H.}\ \bibnamefont {Jen}}, \
  and\ \bibinfo {author} {\bibfnamefont {D.-W.}\ \bibnamefont {Wang}},\ }\href
  {https://journals.aps.org/pra/abstract/10.1103/PhysRevA.90.013631} {\bibfield
   {journal} {\bibinfo  {journal} {Phys. Rev. A}\ }\textbf {\bibinfo {volume}
  {90}},\ \bibinfo {pages} {013631} (\bibinfo {year} {2014})}\BibitemShut
  {NoStop}%
\bibitem [{\citenamefont {Li}\ and\ \citenamefont
  {Sarma}(2015)}]{li2015exotic}%
  \BibitemOpen
  \bibfield  {author} {\bibinfo {author} {\bibfnamefont {X.}~\bibnamefont
  {Li}}\ and\ \bibinfo {author} {\bibfnamefont {S.~D.}\ \bibnamefont {Sarma}},\
  }\href {https://www.nature.com/articles/ncomms8137} {\bibfield  {journal}
  {\bibinfo  {journal} {Nat. Commun.}\ }\textbf {\bibinfo {volume} {6}},\
  \bibinfo {pages} {7137} (\bibinfo {year} {2015})}\BibitemShut {NoStop}%
\bibitem [{\citenamefont {Glaetzle}\ \emph {et~al.}(2014)\citenamefont
  {Glaetzle}, \citenamefont {Dalmonte}, \citenamefont {Nath}, \citenamefont
  {Rousochatzakis}, \citenamefont {Moessner},\ and\ \citenamefont
  {Zoller}}]{PhysRevX.4.041037}%
  \BibitemOpen
  \bibfield  {author} {\bibinfo {author} {\bibfnamefont {A.~W.}\ \bibnamefont
  {Glaetzle}}, \bibinfo {author} {\bibfnamefont {M.}~\bibnamefont {Dalmonte}},
  \bibinfo {author} {\bibfnamefont {R.}~\bibnamefont {Nath}}, \bibinfo {author}
  {\bibfnamefont {I.}~\bibnamefont {Rousochatzakis}}, \bibinfo {author}
  {\bibfnamefont {R.}~\bibnamefont {Moessner}}, \ and\ \bibinfo {author}
  {\bibfnamefont {P.}~\bibnamefont {Zoller}},\ }\href {\doibase
  10.1103/PhysRevX.4.041037} {\bibfield  {journal} {\bibinfo  {journal} {Phys.
  Rev. X}\ }\textbf {\bibinfo {volume} {4}},\ \bibinfo {pages} {041037}
  (\bibinfo {year} {2014})}\BibitemShut {NoStop}%
\bibitem [{\citenamefont {Tanzi}\ \emph {et~al.}(2019)\citenamefont {Tanzi},
  \citenamefont {Lucioni}, \citenamefont {Fam\`a}, \citenamefont {Catani},
  \citenamefont {Fioretti}, \citenamefont {Gabbanini}, \citenamefont {Bisset},
  \citenamefont {Santos},\ and\ \citenamefont {Modugno}}]{tanzi2019prl}%
  \BibitemOpen
  \bibfield  {author} {\bibinfo {author} {\bibfnamefont {L.}~\bibnamefont
  {Tanzi}}, \bibinfo {author} {\bibfnamefont {E.}~\bibnamefont {Lucioni}},
  \bibinfo {author} {\bibfnamefont {F.}~\bibnamefont {Fam\`a}}, \bibinfo
  {author} {\bibfnamefont {J.}~\bibnamefont {Catani}}, \bibinfo {author}
  {\bibfnamefont {A.}~\bibnamefont {Fioretti}}, \bibinfo {author}
  {\bibfnamefont {C.}~\bibnamefont {Gabbanini}}, \bibinfo {author}
  {\bibfnamefont {R.~N.}\ \bibnamefont {Bisset}}, \bibinfo {author}
  {\bibfnamefont {L.}~\bibnamefont {Santos}}, \ and\ \bibinfo {author}
  {\bibfnamefont {G.}~\bibnamefont {Modugno}},\ }\href {\doibase
  10.1103/PhysRevLett.122.130405} {\bibfield  {journal} {\bibinfo  {journal}
  {Phys. Rev. Lett.}\ }\textbf {\bibinfo {volume} {122}},\ \bibinfo {pages}
  {130405} (\bibinfo {year} {2019})}\BibitemShut {NoStop}%
\bibitem [{\citenamefont {B\"ottcher}\ \emph {et~al.}(2019)\citenamefont
  {B\"ottcher}, \citenamefont {Schmidt}, \citenamefont {Wenzel}, \citenamefont
  {Hertkorn}, \citenamefont {Guo}, \citenamefont {Langen},\ and\ \citenamefont
  {Pfau}}]{bottcher2019transient}%
  \BibitemOpen
  \bibfield  {author} {\bibinfo {author} {\bibfnamefont {F.}~\bibnamefont
  {B\"ottcher}}, \bibinfo {author} {\bibfnamefont {J.-N.}\ \bibnamefont
  {Schmidt}}, \bibinfo {author} {\bibfnamefont {M.}~\bibnamefont {Wenzel}},
  \bibinfo {author} {\bibfnamefont {J.}~\bibnamefont {Hertkorn}}, \bibinfo
  {author} {\bibfnamefont {M.}~\bibnamefont {Guo}}, \bibinfo {author}
  {\bibfnamefont {T.}~\bibnamefont {Langen}}, \ and\ \bibinfo {author}
  {\bibfnamefont {T.}~\bibnamefont {Pfau}},\ }\href {\doibase
  10.1103/PhysRevX.9.011051} {\bibfield  {journal} {\bibinfo  {journal} {Phys.
  Rev. X}\ }\textbf {\bibinfo {volume} {9}},\ \bibinfo {pages} {011051}
  (\bibinfo {year} {2019})}\BibitemShut {NoStop}%
\bibitem [{\citenamefont {Chomaz}\ \emph {et~al.}(2019)\citenamefont {Chomaz},
  \citenamefont {Petter}, \citenamefont {Ilzh\"ofer}, \citenamefont {Natale},
  \citenamefont {Trautmann}, \citenamefont {Politi}, \citenamefont
  {Durastante}, \citenamefont {van Bijnen}, \citenamefont {Patscheider},
  \citenamefont {Sohmen}, \citenamefont {Mark},\ and\ \citenamefont
  {Ferlaino}}]{chomaz2019long}%
  \BibitemOpen
  \bibfield  {author} {\bibinfo {author} {\bibfnamefont {L.}~\bibnamefont
  {Chomaz}}, \bibinfo {author} {\bibfnamefont {D.}~\bibnamefont {Petter}},
  \bibinfo {author} {\bibfnamefont {P.}~\bibnamefont {Ilzh\"ofer}}, \bibinfo
  {author} {\bibfnamefont {G.}~\bibnamefont {Natale}}, \bibinfo {author}
  {\bibfnamefont {A.}~\bibnamefont {Trautmann}}, \bibinfo {author}
  {\bibfnamefont {C.}~\bibnamefont {Politi}}, \bibinfo {author} {\bibfnamefont
  {G.}~\bibnamefont {Durastante}}, \bibinfo {author} {\bibfnamefont {R.~M.~W.}\
  \bibnamefont {van Bijnen}}, \bibinfo {author} {\bibfnamefont
  {A.}~\bibnamefont {Patscheider}}, \bibinfo {author} {\bibfnamefont
  {M.}~\bibnamefont {Sohmen}}, \bibinfo {author} {\bibfnamefont {M.~J.}\
  \bibnamefont {Mark}}, \ and\ \bibinfo {author} {\bibfnamefont
  {F.}~\bibnamefont {Ferlaino}},\ }\href {\doibase 10.1103/PhysRevX.9.021012}
  {\bibfield  {journal} {\bibinfo  {journal} {Phys. Rev. X}\ }\textbf {\bibinfo
  {volume} {9}},\ \bibinfo {pages} {021012} (\bibinfo {year}
  {2019})}\BibitemShut {NoStop}%
\bibitem [{\citenamefont {Li}\ \emph {et~al.}(2016)\citenamefont {Li},
  \citenamefont {Hsieh}, \citenamefont {Mou},\ and\ \citenamefont
  {Wang}}]{li2016emergence}%
  \BibitemOpen
  \bibfield  {author} {\bibinfo {author} {\bibfnamefont {W.-H.}\ \bibnamefont
  {Li}}, \bibinfo {author} {\bibfnamefont {T.-C.}\ \bibnamefont {Hsieh}},
  \bibinfo {author} {\bibfnamefont {C.-Y.}\ \bibnamefont {Mou}}, \ and\
  \bibinfo {author} {\bibfnamefont {D.-W.}\ \bibnamefont {Wang}},\ }\href
  {\doibase 10.1103/PhysRevLett.117.035301} {\bibfield  {journal} {\bibinfo
  {journal} {Phys. Rev. Lett.}\ }\textbf {\bibinfo {volume} {117}},\ \bibinfo
  {pages} {035301} (\bibinfo {year} {2016})}\BibitemShut {NoStop}%
\bibitem [{\citenamefont {Khasseh}\ \emph {et~al.}(2017)\citenamefont
  {Khasseh}, \citenamefont {Abedinpour},\ and\ \citenamefont
  {Tanatar}}]{khasseh2017phase}%
  \BibitemOpen
  \bibfield  {author} {\bibinfo {author} {\bibfnamefont {R.}~\bibnamefont
  {Khasseh}}, \bibinfo {author} {\bibfnamefont {S.~H.}\ \bibnamefont
  {Abedinpour}}, \ and\ \bibinfo {author} {\bibfnamefont {B.}~\bibnamefont
  {Tanatar}},\ }\href {\doibase 10.1103/PhysRevA.96.053611} {\bibfield
  {journal} {\bibinfo  {journal} {Phys. Rev. A}\ }\textbf {\bibinfo {volume}
  {96}},\ \bibinfo {pages} {053611} (\bibinfo {year} {2017})}\BibitemShut
  {NoStop}%
\bibitem [{\citenamefont {Kele\ifmmode~\mbox{\c{s}}\else \c{s}\fi{}}\ \emph
  {et~al.}(2020)\citenamefont {Kele\ifmmode~\mbox{\c{s}}\else \c{s}\fi{}},
  \citenamefont {Zhao},\ and\ \citenamefont {Li}}]{keles2019phase}%
  \BibitemOpen
  \bibfield  {author} {\bibinfo {author} {\bibfnamefont {A.}~\bibnamefont
  {Kele\ifmmode~\mbox{\c{s}}\else \c{s}\fi{}}}, \bibinfo {author}
  {\bibfnamefont {E.}~\bibnamefont {Zhao}}, \ and\ \bibinfo {author}
  {\bibfnamefont {X.}~\bibnamefont {Li}},\ }\href {\doibase
  10.1103/PhysRevA.101.023624} {\bibfield  {journal} {\bibinfo  {journal}
  {Phys. Rev. A}\ }\textbf {\bibinfo {volume} {101}},\ \bibinfo {pages}
  {023624} (\bibinfo {year} {2020})}\BibitemShut {NoStop}%
\bibitem [{\citenamefont {Davoudi}\ \emph {et~al.}(2003)\citenamefont
  {Davoudi}, \citenamefont {Asgari}, \citenamefont {Polini},\ and\
  \citenamefont {Tosi}}]{PhysRevB.68.155112}%
  \BibitemOpen
  \bibfield  {author} {\bibinfo {author} {\bibfnamefont {B.}~\bibnamefont
  {Davoudi}}, \bibinfo {author} {\bibfnamefont {R.}~\bibnamefont {Asgari}},
  \bibinfo {author} {\bibfnamefont {M.}~\bibnamefont {Polini}}, \ and\ \bibinfo
  {author} {\bibfnamefont {M.~P.}\ \bibnamefont {Tosi}},\ }\href {\doibase
  10.1103/PhysRevB.68.155112} {\bibfield  {journal} {\bibinfo  {journal} {Phys.
  Rev. B}\ }\textbf {\bibinfo {volume} {68}},\ \bibinfo {pages} {155112}
  (\bibinfo {year} {2003})}\BibitemShut {NoStop}%
\bibitem [{\citenamefont {Asgari}\ \emph {et~al.}(2004)\citenamefont {Asgari},
  \citenamefont {Davoudi},\ and\ \citenamefont {Tosi}}]{ASGARI2004301}%
  \BibitemOpen
  \bibfield  {author} {\bibinfo {author} {\bibfnamefont {R.}~\bibnamefont
  {Asgari}}, \bibinfo {author} {\bibfnamefont {B.}~\bibnamefont {Davoudi}}, \
  and\ \bibinfo {author} {\bibfnamefont {M.}~\bibnamefont {Tosi}},\ }\href
  {\doibase https://doi.org/10.1016/j.ssc.2004.05.029} {\bibfield  {journal}
  {\bibinfo  {journal} {Solid State Commun.}\ }\textbf {\bibinfo {volume}
  {131}},\ \bibinfo {pages} {301 } (\bibinfo {year} {2004})}\BibitemShut
  {NoStop}%
\bibitem [{\citenamefont {Abedinpour}\ \emph {et~al.}(2014)\citenamefont
  {Abedinpour}, \citenamefont {Asgari}, \citenamefont {Tanatar},\ and\
  \citenamefont {Polini}}]{ABEDINPOUR201425}%
  \BibitemOpen
  \bibfield  {author} {\bibinfo {author} {\bibfnamefont {S.~H.}\ \bibnamefont
  {Abedinpour}}, \bibinfo {author} {\bibfnamefont {R.}~\bibnamefont {Asgari}},
  \bibinfo {author} {\bibfnamefont {B.}~\bibnamefont {Tanatar}}, \ and\
  \bibinfo {author} {\bibfnamefont {M.}~\bibnamefont {Polini}},\ }\href
  {\doibase https://doi.org/10.1016/j.aop.2013.10.006} {\bibfield  {journal}
  {\bibinfo  {journal} {Ann. Phys. (NY)}\ }\textbf {\bibinfo {volume} {340}},\
  \bibinfo {pages} {25 } (\bibinfo {year} {2014})}\BibitemShut {NoStop}%
\bibitem [{\citenamefont {B\"ohm}\ \emph {et~al.}()\citenamefont {B\"ohm},
  \citenamefont {Holler}, \citenamefont {Krotscheck},\ and\ \citenamefont
  {Panholzer}}]{PhysRevB.82.224505}%
  \BibitemOpen
  \bibfield  {author} {\bibinfo {author} {\bibfnamefont {H.~M.}\ \bibnamefont
  {B\"ohm}}, \bibinfo {author} {\bibfnamefont {R.}~\bibnamefont {Holler}},
  \bibinfo {author} {\bibfnamefont {E.}~\bibnamefont {Krotscheck}}, \ and\
  \bibinfo {author} {\bibfnamefont {M.}~\bibnamefont {Panholzer}},\ }\href
  {https://link.aps.org/doi/10.1103/PhysRevB.82.224505} {\bibfield  {journal}
  {\bibinfo  {journal} {Phys. Rev. B}\ }\textbf {\bibinfo {volume} {82}},\
  \bibinfo {pages} {224505}}\BibitemShut {NoStop}%
\bibitem [{\citenamefont {Krotscheck}(2000)}]{Krotscheck2000}%
  \BibitemOpen
  \bibfield  {author} {\bibinfo {author} {\bibfnamefont {E.}~\bibnamefont
  {Krotscheck}},\ }\href {https://doi.org/10.1023/A:1004664619961} {\bibfield
  {journal} {\bibinfo  {journal} {Journal of Low Temperature Physics}\ }\textbf
  {\bibinfo {volume} {119}},\ \bibinfo {pages} {103–145} (\bibinfo {year}
  {2000})}\BibitemShut {NoStop}%
\bibitem [{\citenamefont {Kallio}\ and\ \citenamefont
  {Piilo}(1996)}]{kallio1996novel}%
  \BibitemOpen
  \bibfield  {author} {\bibinfo {author} {\bibfnamefont {A.}~\bibnamefont
  {Kallio}}\ and\ \bibinfo {author} {\bibfnamefont {J.}~\bibnamefont {Piilo}},\
  }\href {\doibase 10.1103/PhysRevLett.77.4237} {\bibfield  {journal} {\bibinfo
   {journal} {Phys. Rev. Lett.}\ }\textbf {\bibinfo {volume} {77}},\ \bibinfo
  {pages} {4237} (\bibinfo {year} {1996})}\BibitemShut {NoStop}%
\bibitem [{\citenamefont {Panholzer}\ \emph {et~al.}(2019)\citenamefont
  {Panholzer}, \citenamefont {Hobbiger},\ and\ \citenamefont
  {B{\"o}hm}}]{panholzer2018optimized}%
  \BibitemOpen
  \bibfield  {author} {\bibinfo {author} {\bibfnamefont {M.}~\bibnamefont
  {Panholzer}}, \bibinfo {author} {\bibfnamefont {R.}~\bibnamefont {Hobbiger}},
  \ and\ \bibinfo {author} {\bibfnamefont {H.}~\bibnamefont {B{\"o}hm}},\
  }\href {https://journals.aps.org/prb/abstract/10.1103/PhysRevB.99.195156}
  {\bibfield  {journal} {\bibinfo  {journal} {Phys. Rev. B}\ }\textbf {\bibinfo
  {volume} {99}},\ \bibinfo {pages} {195156} (\bibinfo {year}
  {2019})}\BibitemShut {NoStop}%
\bibitem [{\citenamefont {Giuliani}\ and\ \citenamefont
  {Vignale}()}]{giuliani2005quantum}%
  \BibitemOpen
  \bibfield  {author} {\bibinfo {author} {\bibfnamefont {G.}~\bibnamefont
  {Giuliani}}\ and\ \bibinfo {author} {\bibfnamefont {G.}~\bibnamefont
  {Vignale}},\ }\href@noop {} {\emph {\bibinfo {title} {Quantum Theory of the
  Electron Liquid}}}\BibitemShut {NoStop}%
\bibitem [{\citenamefont {Seydi}\ \emph {et~al.}(2018)\citenamefont {Seydi},
  \citenamefont {Abedinpour}, \citenamefont {Asgari},\ and\ \citenamefont
  {Tanatar}}]{Seydi_PRA2018}%
  \BibitemOpen
  \bibfield  {author} {\bibinfo {author} {\bibfnamefont {I.}~\bibnamefont
  {Seydi}}, \bibinfo {author} {\bibfnamefont {S.~H.}\ \bibnamefont
  {Abedinpour}}, \bibinfo {author} {\bibfnamefont {R.}~\bibnamefont {Asgari}},
  \ and\ \bibinfo {author} {\bibfnamefont {B.}~\bibnamefont {Tanatar}},\ }\href
  {\doibase 10.1103/PhysRevA.98.063623} {\bibfield  {journal} {\bibinfo
  {journal} {Phys. Rev. A}\ }\textbf {\bibinfo {volume} {98}},\ \bibinfo
  {pages} {063623} (\bibinfo {year} {2018})}\BibitemShut {NoStop}%
\bibitem [{\citenamefont {Asgari}\ and\ \citenamefont
  {Tanatar}(2006)}]{asgari2006many}%
  \BibitemOpen
  \bibfield  {author} {\bibinfo {author} {\bibfnamefont {R.}~\bibnamefont
  {Asgari}}\ and\ \bibinfo {author} {\bibfnamefont {B.}~\bibnamefont
  {Tanatar}},\ }\href {\doibase 10.1103/PhysRevB.74.075301} {\bibfield
  {journal} {\bibinfo  {journal} {Phys. Rev. B}\ }\textbf {\bibinfo {volume}
  {74}},\ \bibinfo {pages} {075301} (\bibinfo {year} {2006})}\BibitemShut
  {NoStop}%
\bibitem [{\citenamefont {Gradshteyn}\ \emph {et~al.}(1996)\citenamefont
  {Gradshteyn}, \citenamefont {Jeffrey},\ and\ \citenamefont
  {Ryzhik}}]{gradshteyn1996table}%
  \BibitemOpen
  \bibfield  {author} {\bibinfo {author} {\bibfnamefont {I.}~\bibnamefont
  {Gradshteyn}}, \bibinfo {author} {\bibfnamefont {A.}~\bibnamefont {Jeffrey}},
  \ and\ \bibinfo {author} {\bibfnamefont {I.}~\bibnamefont {Ryzhik}},\ }\href
  {https://books.google.com/books?id=QVaVmgEACAAJ} {\emph {\bibinfo {title}
  {Table of Integrals, Series, and Products}}}\ (\bibinfo  {publisher}
  {Academic Press},\ \bibinfo {year} {1996})\BibitemShut {NoStop}%
\bibitem [{\citenamefont {Singwi}\ \emph {et~al.}(1968)\citenamefont {Singwi},
  \citenamefont {Tosi}, \citenamefont {Land},\ and\ \citenamefont
  {Sj\"olander}}]{PhysRev.176.589}%
  \BibitemOpen
  \bibfield  {author} {\bibinfo {author} {\bibfnamefont {K.~S.}\ \bibnamefont
  {Singwi}}, \bibinfo {author} {\bibfnamefont {M.~P.}\ \bibnamefont {Tosi}},
  \bibinfo {author} {\bibfnamefont {R.~H.}\ \bibnamefont {Land}}, \ and\
  \bibinfo {author} {\bibfnamefont {A.}~\bibnamefont {Sj\"olander}},\ }\href
  {\doibase 10.1103/PhysRev.176.589} {\bibfield  {journal} {\bibinfo  {journal}
  {Phys. Rev.}\ }\textbf {\bibinfo {volume} {176}},\ \bibinfo {pages} {589}
  (\bibinfo {year} {1968})}\BibitemShut {NoStop}%
\bibitem [{\citenamefont {Seydi}(2019)}]{Seydi_phd}%
  \BibitemOpen
  \bibfield  {author} {\bibinfo {author} {\bibfnamefont {I.}~\bibnamefont
  {Seydi}},\ }\emph {\bibinfo {title} {Ground-state and dynamical properties of
  ultra-cold quantum gases with long-range interactions}},\ \href@noop {}
  {Ph.D. thesis},\ \bibinfo  {school} {Institute for Advanced Studies in Basic
  Sciences (IASBS), Zanjan, Iran} (\bibinfo {year} {2019})\BibitemShut
  {NoStop}%
\end{thebibliography}%

\end{document}